\documentclass[preprint2]{aastex631}

\submitjournal{ApJ}

\shorttitle{Galaxy-Cosmic Web Alignments in Subaru PFS}
\shortauthors{Zhang et al.}
\graphicspath{{./}{figures/}}

\usepackage{amsmath}
\usepackage{appendix}
\usepackage[caption=false]{subfig}
\usepackage{savesym}
\savesymbol{tablenum}
\usepackage{siunitx}
\restoresymbol{SIX}{tablenum}
\usepackage[above]{placeins}
\usepackage{makecell}

\usepackage{framed}
\newcommand{\mpc}{\ensuremath{h^{-1}\,\mathrm{cMpc}}}
\newcommand{\mpccubed}{\ensuremath{h^{-3}\,\mathrm{cMpc}^3}}
\newcommand{\invmpccubed}{\ensuremath{h^{3}\,\mathrm{cMpc}^{-3}}}
\newcommand{\zone}{$z=1$}
\newcommand{\ztwo}{$z=2$}
\newcommand{\msun}{M$_\odot$}
\newcommand{\msunh}{\ensuremath{h^{-1}\,M_\odot}}
\newcommand{\mstar}{\ensuremath{M_*}}
\newcommand{\lya}{Ly$\alpha$}
\newcommand{\eone}{\ensuremath{\boldsymbol{\hat{e}_1}}}
\newcommand{\etwo}{\ensuremath{\boldsymbol{\hat{e}_2}}}
\newcommand{\ethree}{\ensuremath{\boldsymbol{\hat{e}_3}}}

\newcommand{\fakespec}{\textsc{fake\_spectra}}
\newcommand{\tardis}{TARDIS}
\newcommand{\alignment}{\ensuremath{\lvert \cos{\theta} \rvert}}
\newcommand{\meanalign}{\ensuremath{\langle \alignment{} \rangle}}
\newcommand{\masshalfrad}{\ensuremath{M_{R/2}}}

\newcommand{\maplotcap}{\meanalign{} summary boxplot for true density field and mock reconstruction stages, for shape alignments. 3D mean alignments are shown on the top row, while projected mean alignments are on the bottom. The null distribution's \meanalign{} is marked with a dashed line for both rows. The boxplot whiskers have been modified to display 16th and 84th percentiles \meanalign{}.}
\newcommand{\sig}{\ensuremath{\sqrt{\chi^2}}}
\newcommand{\zonenormal}{\texttt{Z1\_FID}}
\newcommand{\ztwonormal}{\texttt{Z2\_FID}}
\newcommand{\ztwodouble}{\texttt{Z2\_2X}}

\newcommand{\figwidth}{16cm}
\newcommand{\legendinclude}[1]{\hspace{3em}\includegraphics[width=\textwidth]{#1}}
\newcommand{\mainsigfigs}{2}

\let\oldequation\equation
\let\oldendequation\endequation

\renewenvironment{equation}
  {\linenomathNonumbers\oldequation}
  {\oldendequation\endlinenomath}

\begin{document}

\title{Alignments Between Galaxies and the Cosmic Web at $z\sim 1-2$ in the IllustrisTNG Simulations}

\author[0000-0003-1465-135X]{Benjamin Zhang}
\affiliation{Department of Physics and Astronomy, University of Southern California, Los Angeles, CA 90089, USA}
\email{zhangben@usc.edu}

\author[0000-0001-9299-5719]{Khee-Gan Lee}
\affiliation{Kavli IPMU (WPI), UTIAS, The University of Tokyo, Kashiwa, Chiba 277-8583, Japan}

\author[0000-0003-2183-7021]{Alex Krolewski}
\affiliation{Perimeter Institute for Theoretical Physics, 31 Caroline St. North, Waterloo, ON N2L 2Y5, Canada}
\affiliation{Waterloo Centre for Astrophysics, University of Waterloo, Waterloo, ON N2L 3G1, Canada}

\author[0000-0001-9879-4926]{Jingjing Shi}
\affiliation{Kavli IPMU (WPI), UTIAS, The University of Tokyo, Kashiwa, Chiba 277-8583, Japan}

\author[0000-0001-7832-5372]{Benjamin Horowitz}
\affiliation{Princeton Department of Astrophysical Sciences, Princeton, NJ 08544, USA}

\author[0000-0002-1008-6675]{Robin Kooistra}
\affiliation{Kavli IPMU (WPI), UTIAS, The University of Tokyo, Kashiwa, Chiba 277-8583, Japan}

\begin{abstract}

    Galaxy formation theories predict that galaxy shapes and angular momenta have non-random alignments with the cosmic web. This leads to so-called intrinsic alignment between pairs of galaxies, which is important to quantify as a nuisance parameter for weak lensing. We study galaxy-cosmic web alignment in the IllustrisTNG suite of hydrodynamical simulations at redshifts 1 and 2, finding that alignment trends are consistent with previous studies. However, we find that the magnitude of the spin alignment signal is $\sim 2.4 \times$ weaker than seen in previous studies of the Horizon-AGN simulation, suggesting that this signal may have significant dependence on subgrid physics. Based on IllustrisTNG, we then construct mock observational spectroscopic surveys that can probe shape-cosmic web alignment at $z \sim 1-2$, modeled on the low-$z$ galaxy redshift and IGM tomography surveys on the upcoming Subaru Prime Focus Spectrograph Galaxy Evolution (PFS GE) survey. However, even over box sizes of $L=205\,h^{-1}\,\mathrm{Mpc}$, we find that global anisotropies induce a sample variance in the 2D projected alignment signal that depend on the projected direction --- this induces significant errors in the observed alignment. We predict a $5.3\sigma$ detection of IllustrisTNG's shape alignment signal at $z \sim 1$ from Subaru PFS GE, although a detection would be challenging at $z \sim 2$. However, a rough rescaling of the relative alignment signal strengths between the TNG and HorizonAGN simulations suggests that PFS GE should be able to more easily constrain the latter's stronger signal.

\end{abstract}


\section{Introduction} \label{sec:intro}

    In the prevailing paradigm of inflationary Big Bang cosmology, tiny overdensities
    in the initial Gaussian density field grow via gravity, eventually reaching a point
    at which they decouple from the Hubble flow. Collapsing under their own gravity, virialized structures referred to as
    dark matter halos \citep{GunnGott72} grow via accretion
    from the surrounding large-scale structure:
     smaller halos assemble before merging into larger
    ones in a hierarchical fashion
    \citep{Blumenthal84,Wechsler02}.
    As dark matter collapses around
    halos, on larger scales ($\gtrsim$ Mpc) it forms a distinctive ``cosmic web,''
    with filaments connecting halos and sheets surrounding the large empty voids
    \citep{zel82,kly+shand83,ein84,defw85,geller+huchra89,bond+96}.
    The cosmic web is intimately related
    to the dynamics of dark matter:
    gravitational collapse first occurs onto the plane of sheets, then radially perpendicular to the long axis of filaments, and finally along all three directions into nodes or halos \citep{Zeldovich1970}.  Matter thus moves out of voids, along filaments and sheets, and ultimately into halos.
    
    Halo shapes and angular momenta
    are determined by these accretion processes, linking
    halo properties to the properties
    of the surrounding dark matter.
    On small ($\sim$ Mpc) scales, the connection
    between the cosmic web and halo spins and shapes is in principle a sensitive probe of halo formation and evolution.
    This connection also gives rise to so-called intrinsic alignments between pairs of observed galaxies, a
    significant systematic in weak lensing surveys \citep{kirk_observation_review,kiessling_theory_review} which can bias cosmological inference if not properly modelled  \citep{bridle+07,kirk+12,Krause16,Blazek19}.

    Simulations find that the major axis of the halo inertia
    tensor tends to be preferentially aligned along filaments
    \citep{alt+06,hahn+07_evolution,ac+07,zhang+09,LibeskindAligns,fr+14}
     and parallel to the surface of voids \citep{pat+06,brun+07,cuesta+08}.
    Parallel alignment is generally found across all halo
    masses, though it is typically weaker at low masses
    and stronger at high masses \citep{hahn+07_evolution,ac+07,zhang+09,LibeskindAligns,fr+14,chen+15,Codis2015Spin,gan18,Pandya}.
    
    Theory further predicts that halo spin alignments transition from preferentially
    parallel to filaments at low mass, to perpendicular at high mass.
    At low mass, halos grow by accretion of matter in the plane perpendicular to the filament, creating a rotating disk in this plane \citep{pich+11,codis+12}. High mass halos grow primarily by mergers, converting motion along the filament into angular momentum. This leads to a characteristic transition mass at which halos flip from aligned to anti-aligned, typically $M_{\rm halo} \sim 10^{12} M_{\odot}$ \citep{bs+05,ac07,hahn+07_evolution,codis+12,trow+13,ac14,gan18,wang18,Lopez19,gan21}.
    The alignment signal appears to depend on redshift: \citet{codis+12} find that the transition mass for halo spins declines with time to higher masses.
    
    Of course, strong alignment between the cosmic web and dark matter halos does not necessarily imply the same degree of alignment between the cosmic web and baryonic galaxies. 
    However, simulations suggest that during the main epoch of galaxy formation (at high redshift), the shape of baryonic matter tends to trace the halo shape. 
    This weakens during later times, but is still significant at low redshift \citep{TomassettiHaloGalMisalign}. 
    In addition, while interactions between group centrals and satellites are expected to decrease alignments between the cosmic web and galaxies, observed satellite fractions at $z \sim 1-2$ are $\lesssim 0.1$, so that the effects of central-satellite galaxy alignments are relatively minor at higher redshift \citep{mccracken+15, ishikawa+17}. Looking only at cosmic web-galaxy alignment, we therefore expect qualitatively similar alignments to the cosmic web-halo case, which decline in strength at lower redshifts. This is borne out by \citet{Codis2018}, who finds that low-mass galaxies are more strongly aligned at high redshift. \citet{gan19} and \citet{Bate20} also find that high-mass galaxies show stronger perpendicular alignment at high redshift. 
    As with halos, galaxy spins undergo a flip from alignment to anti-alignment at $M_{\star} \sim 10^{10.5} M_{\odot}$ \citep{Horizon-AGN, Codis2015Spin, Codis2018, wang18b, Kraljic20}. However, in contrast with halos, \citet{Codis2018} and \citet{Kraljic20} do not detect significant redshift evolution in the galaxy spin transition mass.
    
    Observational evidence for this picture is difficult, 
    owing to the weakness of the signal and the high completeness and galaxy density required to reconstruct
    the cosmic web. A number of low-redshift studies have found evidence
    supporting the theoretical picture, in which spiral galaxies
    spin parallel to filaments and elliptical
    galaxies spin perpendicular to filaments \citep{temp13,tl13,pah+16,bb20}, although the significance is weak and some studies do not detect this purported alignment \citep{Krolewski19}. Similarly, \citet{zhang+13}
    find stronger shape alignments for red galaxies
    than for blue galaxies. Meanwhile \citet{Welker20} have presented the first confident detection of the mass-dependent spin flip using filaments from GAMA and galaxy spins from SAMI, and \citet{Kraljic21} report a similar detection using MaNGA spins.
    
    The observational requirements for spin and shape alignment measurements are sufficiently challenging that few studies have pushed beyond $z \sim 0$.
    The main difficulty is a lack of a sufficiently high density spectroscopic surveys to resolve the cosmic web over large volumes, although VIPERS has detected the cosmic web to $z \sim 0.7$ \citep{guzzo,malavasi16}.
    At $z \sim 2$, the technique of Lyman-$\alpha$ forest tomography offers an alternative method to mapping the density field on Mpc scales \citep{pichon2001,cau+08}. 
    As Lyman-$\alpha$ forest spectra are fundamentally one-dimensional tracers of the underlying density, density reconstruction over large volumes only requires dense areal sampling, as opposed to the dense volumetric sampling required for reconstruction from traditional galaxy surveys \citep{lee_obs_req}.
    The CLAMATO Ly$\alpha$ tomographic survey \citep{CLAMATO_DR1,CLAMATO_DR2} has reconstructed the $z \sim 2$ density field over a volume of $\num{4.1e5}$  $\mpccubed{}$ in the COSMOS field to a resolution of $2 \sim 3$ Mpc. However, with only 810 unique coeval galaxies across existing spectrographic surveys in the reconstructed volume, it is unable to conclusively detect alignments \citep{krolewski:2017, momose+22}.
    The upcoming Subaru Prime Focus Spectrograph Galaxy Evolution Survey offers an improvement on both fronts:
    a high-density spectroscopic survey \citep{pfs-whitepaper} that can resolve the cosmic web at $z \sim 1-2$ with $N_g \sim$250,000 galaxies, and a Ly$\alpha$ forest tomography survey $z \sim 2.5$ that will also cover $N_g\sim$ 20,000 coeval galaxies. In this paper, we explore the prospects for constraining galaxy alignments with upcoming Subaru PFS data.
    
    In Section \ref{sec:sim}, we first explore idealized galaxy alignment mass trends from the IllustrisTNG cosmological simulation, with an eye towards estimating the galaxy spin transition mass and constraining the strength of alignment expected from observational surveys. 
    We describe the IllustrisTNG data used in Section \ref{sec:sim:methods}, and discuss our results in Section \ref{sec:sim:results}.
    In Section \ref{sec:mock:methods}, we construct a mock Subaru PFS observational survey from the IllustrisTNG simulation, with appropriate galaxy selections and an accompanying Ly$\alpha$ forest tomography survey.
    Finally, we present and discuss our results for the alignment constraints produced by our mock survey in Section \ref{sec:mock:results}.
    
    Throughout this work, we adopt the \cite{Planck2016} concordance $\Lambda$CDM cosmology ($\Omega_m$ = 0.3, $\Omega_b$ = 0.047, $h$ = 0.685, $n_s$ = 0.965, $\sigma_8$ = 0.8).

\section{Simulation Results}
\label{sec:sim}

    \subsection{Data and Methods}
    \label{sec:sim:methods}

        \subsubsection{IllustrisTNG simulations}

            In this work, we use the IllustrisTNG suite of cosmological magnetohydrodynamic simulations, which is based on the Arepo adaptive mesh refinement code. For our main analysis we use the TNG300-1 simulation box, which has $2500^3$ dark matter particles, and a side length $L=205\,\mpc{}$, for a total volume of \num{8.61e6} \mpccubed{} \citep{tng-0, tng-1, tng-2, tng-3, tng-4, tng-5}. This offers the largest cosmological volume and galaxy statistics in the publicly available IllustrisTNG suite, and is better matched to upcoming spectroscopic surveys than its higher-resolution counterparts.
            We will also use the TNG100-1 volume to compare the effect of simulation grid resolution on our shape and spin alignment trends. This higher-resolution suite has 1820$^3$ dark matter particles with a side length of $L=75$ $\mpc{}$.
            Of the $100$ cosmological-redshift snapshots of the simulations that span $z=127$ to $z=0$, we use the data products at redshifts \ztwo{} and \zone{} for this work.
            While this does not exactly match the expected redshift ranges of $z \sim 2.2-2.7$ and $z \sim 0.7-1.7$ for which the Subaru PFS Galaxy Evolution Survey will obtain galaxy spectra, we believe the two selected redshifts are representative and our conclusion should still hold.
            The baryon particles in TNG300-1/TNG100-1 have mass resolution of $\num{7.6e6}$ \msun{}/$\num{9.4e5}$ \msun{} respectively, while the dark matter particle masses are $\num{5.9e7}$ \msun{}/$\num{7.5e6}$ \msun{} respectively. These resolutions are  sufficient to resolve both the large-scale morphologies of galaxies and the surrounding cosmic web in detail.
            
            The IllustrisTNG suite assumes the same \citet{Planck2016} cosmology we use in this work.
            To construct the matter density field, we construct a $512^3$ grid mesh and assign each dark matter particle to its nearest cell.
            
        \subsubsection{Galaxy properties}
        \label{sec:sim:methods:shapes}
        
            In IllustrisTNG, subhalos are identified via the \textsc{subfind} algorithm developed by \citet{subfind}. %
            For our analysis of morphological alignment trends in TNG300-1 and TNG100-1, we use the morphology calculated by \citet{jingjing-shapes}. This morphology sample is calculated for subhalos which have a cosmological origin (\textit{SubhaloFlag $\neq$ 0}), and have $\masshalfrad{} > 10^9$ \msunh{}, where \masshalfrad{} is the stellar mass within twice the half-stellar-mass radius. However, we note that for the rest of this work, we use ``stellar mass" to refer to total stellar mass to be consistent with previous works, such as \citet{Codis2018}. 
            Our sample includes both galaxies classified as central galaxies (most massive within each dark matter halo) and satellite galaxies (all others associated with a given dark matter halo).
            
            To measure galaxy morphology, \citet{jingjing-shapes} adopts the reduced mass inertia tensor formalism described in \citet{reduced-mass-inertia-tensor}.
            In this formalism, galaxies are modelled as 3D triaxial ellipsoids, with axis lengths as $a \geq b \geq c$, and corresponding axis vectors as $\vec{a}, \vec{b}, \vec{c}$.
            These quantities are found by calculating the sorted eigenvalues and eigenvectors respectively of the reduced mass inertia tensor $\Tilde{I}$:
            \begin{equation}
                \Tilde{I}_{ij} = \frac{\sum_n m_n \frac{x_{ni}x_{nj}}{r^2_n}}{\sum_n m_n}
            \end{equation}
            where $0 \leq i, j \leq 2$ represent coordinate indices, $m_n$ is the mass of the $n^{th}$ particle in the galaxy, and $x_{ni}$ represents the $i^{th}$ coordinate of the $n^{th}$ particle (with respect to the galaxy center of mass).
            $r_n^2$ is defined as:
            \begin{equation*}
                r_n^2 = \sum_i x^2_{ni}
            \end{equation*}
            We henceforth refer to the longest axis $\vec{a}$ as the `shape' direction of each galaxy in this sample.
            
            For our analysis of spin alignment trends in TNG300-1 and TNG100-1, we will use the stellar particle-only spin also calculated by \cite{jingjing-shapes} which we refer to as the `spin' sample.
            We limit our sample to subhalos with more than 50 stellar particles, which ensures that they have a well-defined spin vector following the convention set by \citet{Codis2018}.
            
            Table \ref{table:gal-number-mass} summarizes the number of galaxies in each simulation and redshift sample, and gives 5\% and 95\% total stellar mass percentiles for each sample.
            
            \begin{table*}[ht!]
            \begin{tabular}{lccccc}
            \hline
            \textbf{Simulation} & \textbf{Redshift} & \textbf{Sample} & \textbf{\# galaxies} & \textbf{5\%ile \mstar{} (\msunh{})} & \textbf{95\%ile \mstar{} (\msunh{})} \\ \hline
            TNG300-1            & z=2               & Shape           & 128,019              & \num{1.348882e+09}              & \num{3.827169e+10}               \\ \cline{3-6} 
                                &                   & Spin            & 256,309              & \num{3.735926e+08}              & \num{2.392023e+10}               \\ \cline{2-6} 
                                & z=1               & Shape           & 194,189              & \num{1.384419e+09}              & \num{4.568499e+10}               \\ \cline{3-6} 
                                &                   & Spin            & 345,296              & \num{3.731775e+08}              & \num{3.260130e+10}               \\ \hline
            TNG100-1            & z=2               & Shape           & 8,530                & \num{1.409582e+09}              & \num{3.984801e+10}               \\ \cline{3-6}
                                &                   & Spin            & 61,976               & \num{4.677695e+07}              & \num{6.630453e+09}               \\ \cline{2-6} 
                                & z=1               & Shape           & 13,245               & \num{1.394461e+09}              & \num{5.170797e+10}               \\ \cline{3-6}
                                &                   & Spin            & 65,641               & \num{4.503330e+07}              & \num{1.266712e+10}               \\ \hline
            \end{tabular}
                \caption{Number of galaxies in each sample, with 5\% and 95\% percentiles of the total stellar mass. Shape samples have reduced inertia tensor morphologies calculated by \citet{jingjing-shapes}, and are selected via a stellar mass threshold, while spin samples only have angular momentum vectors, and are selected via a stellar particle number threshold.}
                \label{table:gal-number-mass}
            \end{table*}
            
        \subsubsection{Deformation tensor and alignment angle}
        \label{sec:sim:methods:align}
            In order to quantify the large-scale dynamics of the matter density field without directly using peculiar velocities, we adopt the deformation tensor approach of \cite{krolewski:2017} and \cite{lee_white+16}, which in turn is based on the basic formalism described in \cite{HahnDeformationTensor} and \citet{ForeroRomero2009TWeb}.
            The deformation tensor is defined as the Hessian of the gravitational potential $\phi$ of each point in space:
            \begin{equation}
                D_{ij} = \partial x_i \partial x_j \phi
            \end{equation}
            In practice, the deformation tensor is efficiently calculated in Fourier space directly from the density, as described in \cite{krolewski:2017}.
            
            Diagonalizing the tensor yields a set of 3 eigenvalues and associated normalized eigenvectors for each point.
            By convention, we sort the eigenvalues such that $\lambda_1 > \lambda_2 > \lambda_3$.
            Whether the eigenvalues are indexed in increasing or decreasing order varies from work to work, but produces the same results with the indices switched accordingly.
            In the Zel'dovich approximation, the deformation tensor solely determines the time evolution of the initial density field \citep{Zeldovich1970}:
            matter first collapses along the eigenvector $\eone{}$ associated with $\lambda_1$, and collapses last along $\ethree{}$.
            Within the approximation, a positive/negative eigenvalue indicates that matter is collapsing/expanding along the axis of the associated eigenvector.
            
            The dark matter density is first smoothed to eliminate the effect of highly nonlinear fluctuations in density before calculating the deformation tensor. We apply a Gaussian kernel with standard deviation 2 $\mpc{}$, which is comparable to smoothing scales used by other shape and spin alignment studies \citep{Codis2015Spin, fr+14}.
            
            In order to characterize the degree to which a galaxy's shape or spin vector is aligned with the underlying matter deformation tensor at the galaxy's position, we calculate \alignment{}, where $\theta$ is the angle between the shape/spin and each of the deformation tensor's eigenvectors at the nearest density grid point to the galaxy.
            We take the absolute value since the sign convention of the eigenvectors is arbitrary. $\alignment{}>0.5$ means the two vectors are largely parallel with each other, while $\alignment{} < 0.5$ means they tend to be perpendicular with each other.
            
    \subsection{Idealized Simulation Alignments}
    \label{sec:sim:results}
            In this section, we discuss the alignment trends of the simulated galaxy shapes and spins, with respect to the cosmic web that traces the underlying matter density field. 
        \subsubsection{Alignment of galaxy shapes}
        \label{sec:sim:results:shape}
    
            \begin{figure*}
                \centering
                \begin{framed}
                \legendinclude{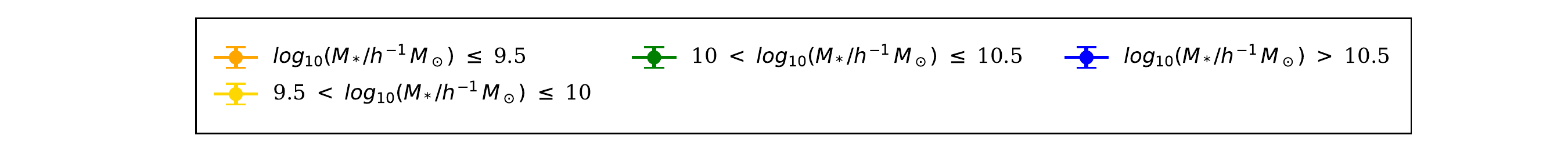}
                \vspace{-2em}
                
                \subfloat[\zone{}]{%
                    \begin{minipage}{0.35\textwidth}
                        \centering
                        \includegraphics[width=\textwidth]{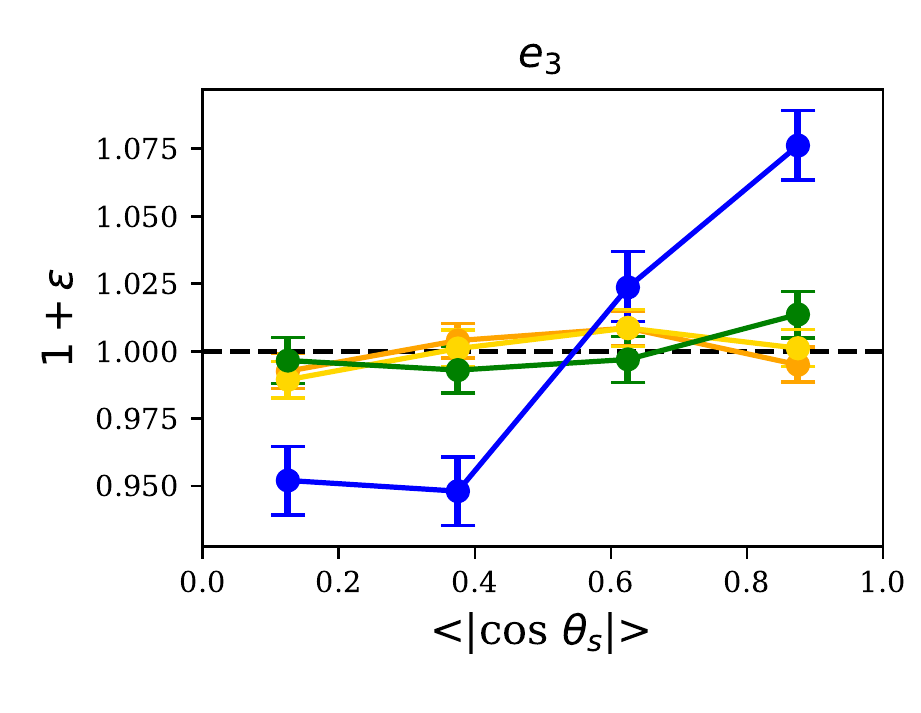}\\
                        \vspace{-1em}
                        \includegraphics[width=\textwidth]{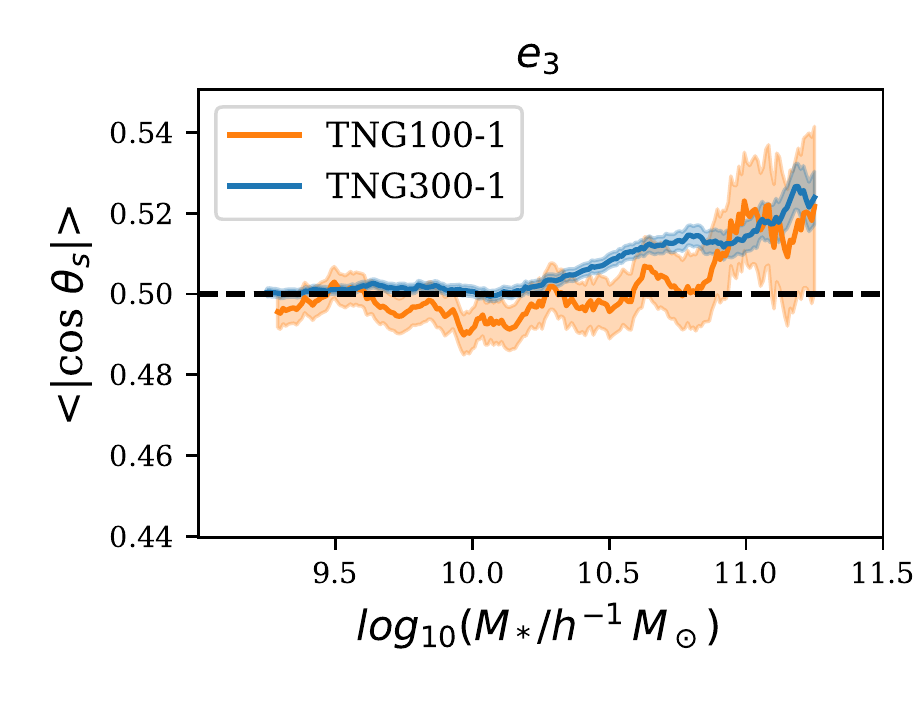}\\
                    \end{minipage}%
                    \label{fig:z1-shape-ideal}
                }%
                \subfloat[\ztwo{}]{%
                    \begin{minipage}{0.35\textwidth}
                        \centering
                        \includegraphics[width=\textwidth]{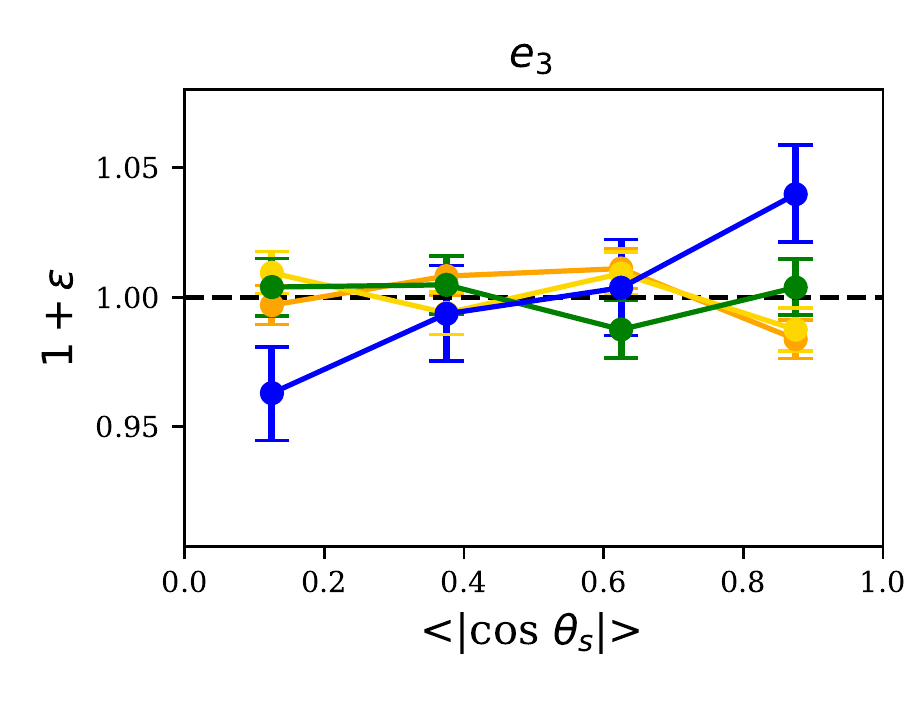}\\
                        \vspace{-1em}
                        \includegraphics[width=\textwidth]{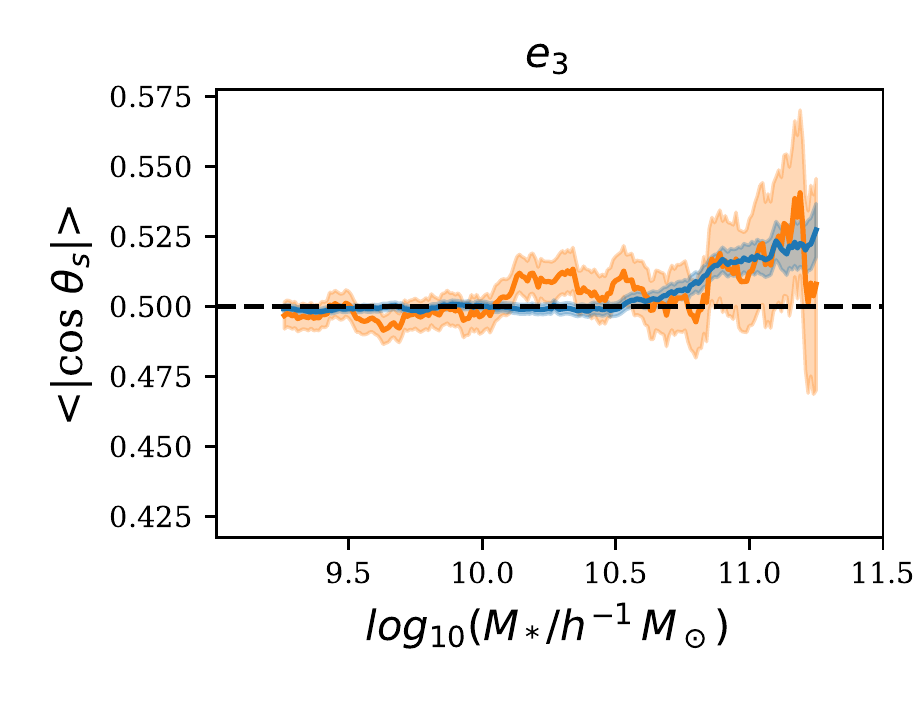}\\
                    \end{minipage}
                    \label{fig:z2-shape-ideal}
                }%
                \end{framed}
                \caption{Shape-cosmic web alignment signal for all galaxies with radius-limited stellar mass $\mstar> 10^{10}\, \msunh$, as a function of total stellar mass. For each redshift, the top panel shows the alignment signal calculated in total stellar mass bins. In the bottom panel, the alignment mean is calculated as a function of a sliding stellar mass bin. Error bars are calculated using bootstrap sampling, with error bars/shaded regions indicating the 16\%-84\% percentile ranges. The subscript on $\theta_s$ indicates that we are calculating shape alignment.}
                \label{fig:all-shape-ideal}
            \end{figure*}

            In Figure \ref{fig:all-shape-ideal}, we show the 3-dimensional alignment signal between shape vectors and the cosmic web for the full TNG300-1 galaxy sample from Section \ref{sec:sim:methods:shapes}, for \zone{} and \ztwo{} respectively.
            As shape-$\etwo{}$ and shape-$\eone{}$ alignments are largely the inverse of shape-$\ethree{}$ alignments (i.e. perpendicular when the latter is parallel), we opt to plot only shape-$\ethree{}$ alignments for the sake of brevity.
            In the top panel of each figure, we compare the distribution of alignment angle \alignment{} in four stellar mass bins.
            We refer to a positively-sloped bin trend (i.e.\ toward an excess at positive \alignment{} to the two vectors) as a positive alignment, and vice versa as a negative alignment (i.e. perpendicular).
            
            On the bottom panel of each figure, we adopt a ``sliding-window" mass binning approach for characterizing the overall alignment trend as a function of stellar mass.
            From all galaxies within a mass range of width $0.5$ in log$_{10}$\mstar{} space, we compute the mean alignment \meanalign{}.
            A mean alignment of $\meanalign = 0.5$ corresponds to random orientations and no systematic alignment.
            In this panel we also show results from TNG100-1, which has a higher particle/cell mass resolution than TNG300-1. The trade off, however, is its reduced volume compared to TNG300-1 (75$^3$ \mpccubed{} vs. 205$^3$ \mpccubed{}, i.e.\ $\sim 20\times$ smaller) which leads to greater cosmic variance in the simulation box. This is reflected in the larger bootstrap error bars.
            The number of galaxies in the highest-mass bin (smallest galaxy number across all bins) is 186 for TNG100-1, so we believe bootstrap errors should still be fairly accurate for the high-mass end.
            
            The \ethree{}-shape alignment shows a clear transition mass at both redshifts: at lower stellar masses there is a flat trend consistent with no alignment, which changes to an increasing alignment trend with stellar mass beyond the transition.
            This is consistent with past studies of dark matter halo shape alignments with \ethree{} \citep{LibeskindAligns, zhang+09}, which are similar to galaxy shape alignments due to central galaxies being mostly aligned with their host halo at high redshifts \citep{TomassettiHaloGalMisalign}.
            The transition mass (the bin center at which the bin's mean alignment is 0.5) is $\mstar \sim 10^{10.5}$ \msunh{} for \ztwo{}, and $\mstar \sim 10^{10.1}$ \msunh{} for \zone{}.
            
            The mass-alignment trends for TNG100-1 are broadly consistent with TNG300-1 (within 1$\sigma$ errorbars for TNG100-1) for \ethree{} at both redshifts.

        \subsubsection{Alignment of galaxy spins}
        \label{sec:sim:results:spin}
        
            \begin{figure*}
                \centering
                \begin{framed}
                \legendinclude{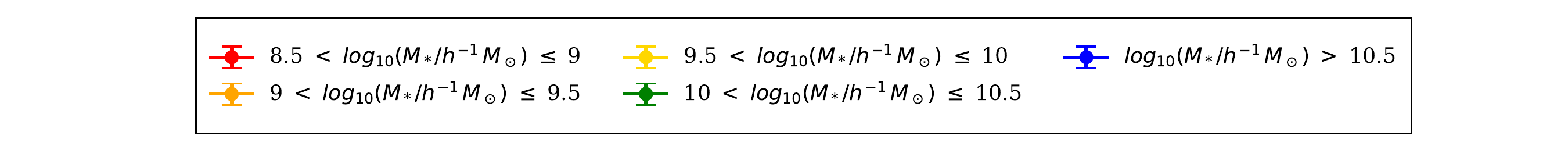}
                \vspace{-2em}
                
                \subfloat[\zone{}]{%
                    \begin{minipage}{0.35\textwidth}
                        \centering
                        \includegraphics[width=\textwidth]{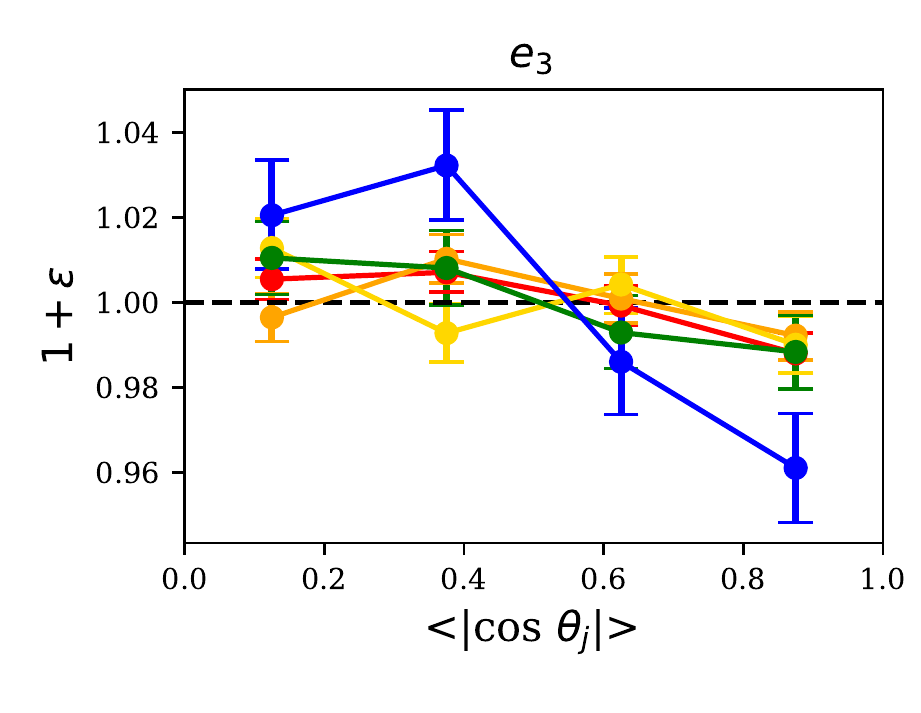}\\
                        \vspace{-1em}
                        \includegraphics[width=\textwidth]{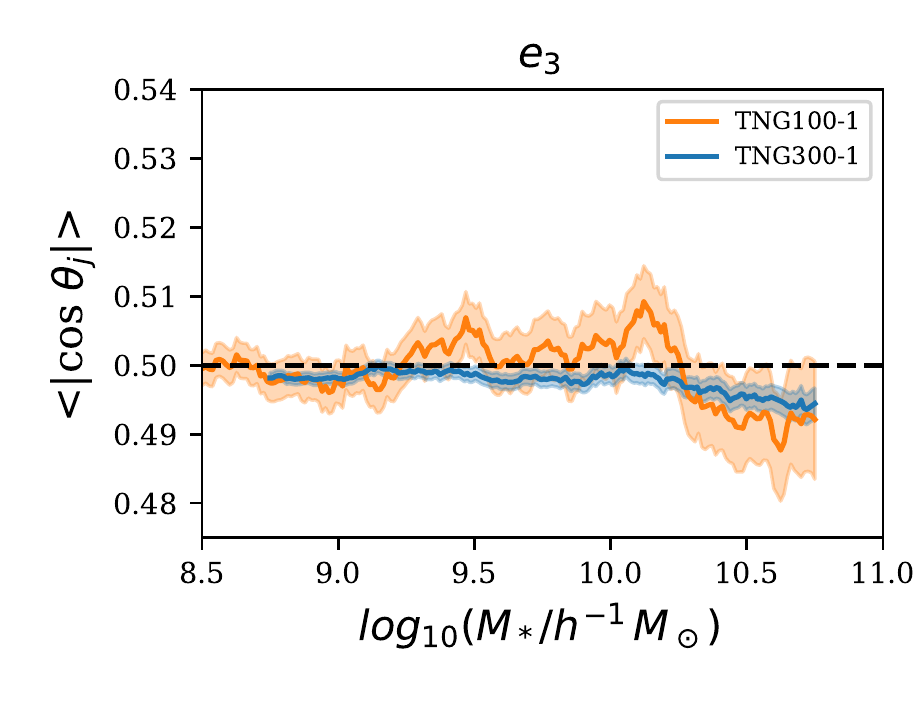}
                    \end{minipage}
                    \label{fig:z1-spin-ideal}
                }%
                \subfloat[\ztwo{}]{%
                    \begin{minipage}{0.35\textwidth}
                        \centering
                        \includegraphics[width=\textwidth]{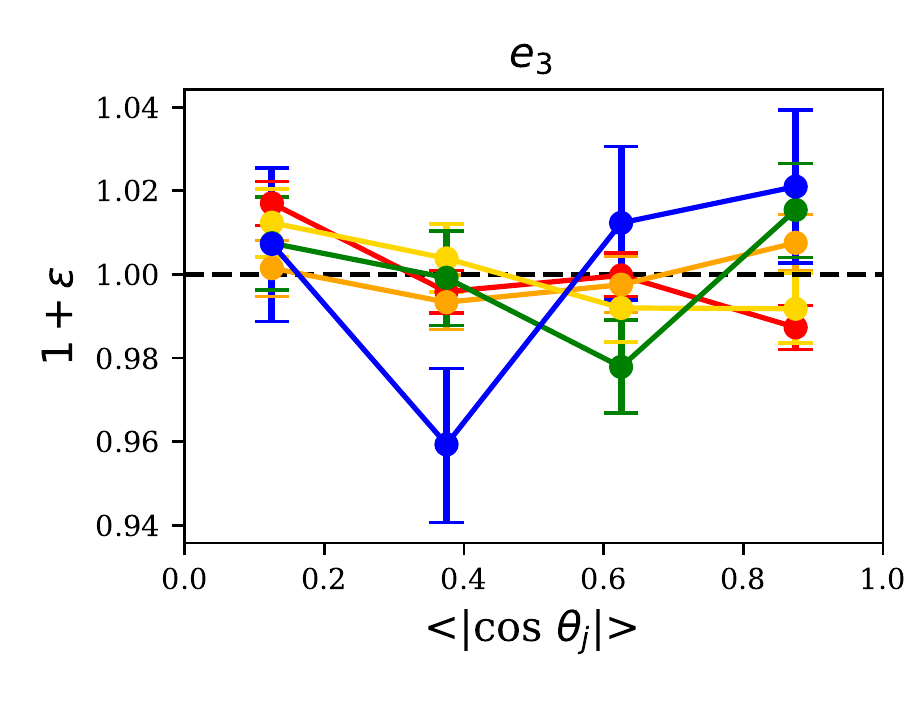}\\
                        \vspace{-1em}
                        \includegraphics[width=\textwidth]{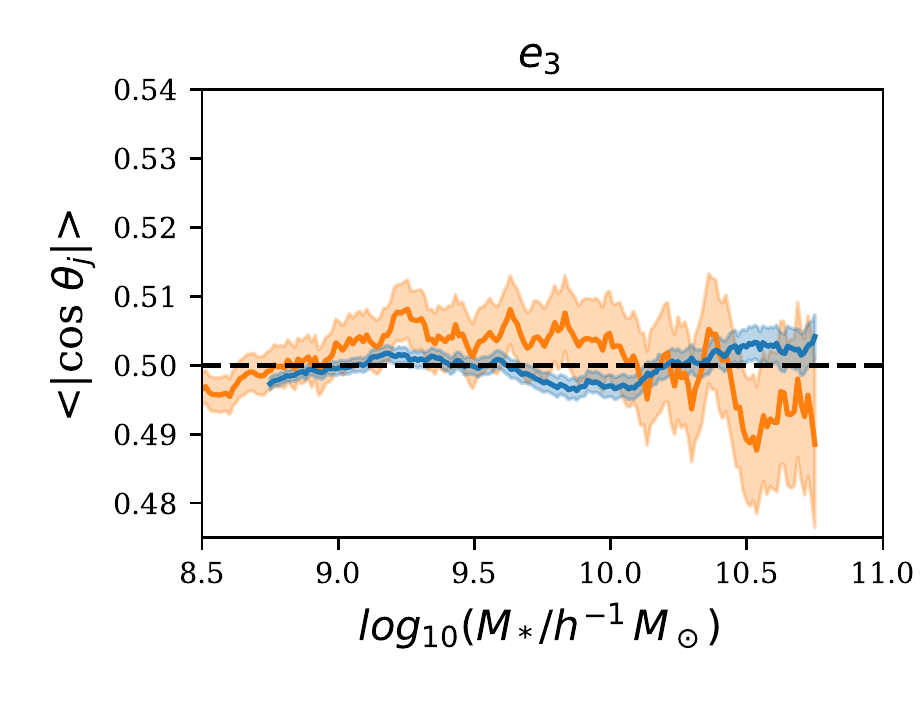}
                    \end{minipage}
                    \label{fig:z2-spin-ideal}
                }%
                \end{framed}
                \caption{Spin-cosmic web alignment signal for all galaxies with more than $50$ stellar particles, as a function of total stellar mass. For each redshift, the top panel shows the detailed alignment signal calculated in total stellar mass bins. The alignment mean is calculated as a function of a sliding stellar mass bin in the below panel. The error bars are calculated using bootstrap sampling, with error bars/shaded regions indicating the 16\%-84\% percentile ranges. The subscript on $\theta_j$ indicates that we are calculating spin alignment.}
                \label{fig:all-spin-ideal}
            \end{figure*}

            In Figure \ref{fig:all-spin-ideal}, we show the alignments between galaxy spins (i.e.\ angular momenta of their stellar content) with respect to the $\ethree{}$ cosmic web eigenvector for varying stellar mass at \zone{} and \ztwo{} respectively.
            As our galaxy sample for spin is larger than that of shape at the low-mass end due to relaxed selection criteria, we show sliding-window bins down to a bin centered at $M_* = 10^{8.5}$ $\msunh{}$, in contrast to the lowest-mass bin of $M_* = 10^9$ $\msunh{}$ for shape alignment.

            At \ztwo{}, we appear to see no trends for spin-\ethree{} alignment at the \textgreater$1\sigma$ level for both TNG100-1 and TNG300-1.
            This is somewhat surprising, as previous DM halo alignment studies \citep{ZhangDMSpin2009, LibeskindAligns} and hydrodynamic galaxy alignment studies \citep{WangIllustrisSpinAlign, Codis2018, Codis2015Spin, Kraljic20} have observed a clear ``spin-flip" transition at stellar masses of $M_{\text{flip}} \sim 10^{9.5} - 10^{10.5}$ \msunh{} akin to the shape alignment transition we observe at \ztwo{}.
            
            At \zone{}, the spin-\ethree{} alignment trend for TNG300-1 has a significant (\textgreater$1\sigma$) transition towards negative alignment at $\mstar =  10^{10.2}$ \msunh{}.
            This mass-dependent behavior has not been seen in past simulated galaxy spin alignment studies such as \citet{WangIllustrisSpinAlign}, who studied the mass dependence of galaxy spin alignment in the predecessor Illustris simulation and report a monotonically decreasing spin-\ethree{} alignment for both \zone{} and \ztwo{}.
            While this difference may come from differing simulation physics and differences in galaxy selection criteria, it may also be partly a result of the broad (order-of-magnitude in \mstar{}) and non-overlapping mass bins used by \citet{WangIllustrisSpinAlign}, which could obscure finer mass trends for alignment.
            
            For both redshifts, TNG100-1 is again broadly consistent with TNG300-1 (within 1$\sigma$ errorbars for TNG100-1).
            
        \subsubsection{Comparison with Horizon-AGN results}
        \label{sec:sim:results:horizon}
        
            To investigate the effects of the implemented sub-grid physics models in cosmological hydro-dynamical simulations on galaxy-cosmic web alignments, in this subsection, we compare the spin alignments of IllustrisTNG with that from the Horizon-AGN simulation.
            The Horizon-AGN simulation \citep{Horizon-AGN} is a hydrodynamic adaptive mesh simulation over a box size of $L=100\, \mpc$, with a dark matter particle mass resolution of \num[]{8e7} \msun{}, which is comparable to the mass resolution of TNG300-1 (\num[]{5.9e7} \msun{}).
            As spin-cosmic web alignments in Horizon-AGN have been extensively studied \citep[e.g.]{Horizon-AGN, Codis2015Spin}, we opt to use it for comparison.
            In particular, we compare the spin alignments from the IllustrisTNG \zone{} with the spin alignments calculated by \citet{Codis2015Spin} for $z \sim 1.2$.
            
            We impose a minimum stellar mass cut on the galaxy sample for TNG300-1 and TNG100-1 used in the previous sections of $M_* \geq 10^{8.06}$ \msunh{}, to match the lower stellar mass limit used in \citet{Codis2015Spin}'s analysis. The TNG300-1 sample is unchanged due to the minimum stellar particle count imposed, but the TNG100-1 sample is reduced by $\sim$37\%.
            We then calculate the spin-cosmic web alignment using the density field smoothed with a  kernel size of $200$ $h^{-1}$ ckpc used in \citet{Codis2015Spin}, instead of the 2 \mpc{} kernel used in the rest of our work.

            \begin{figure*}
                \centering
                \includegraphics[width=\figwidth]{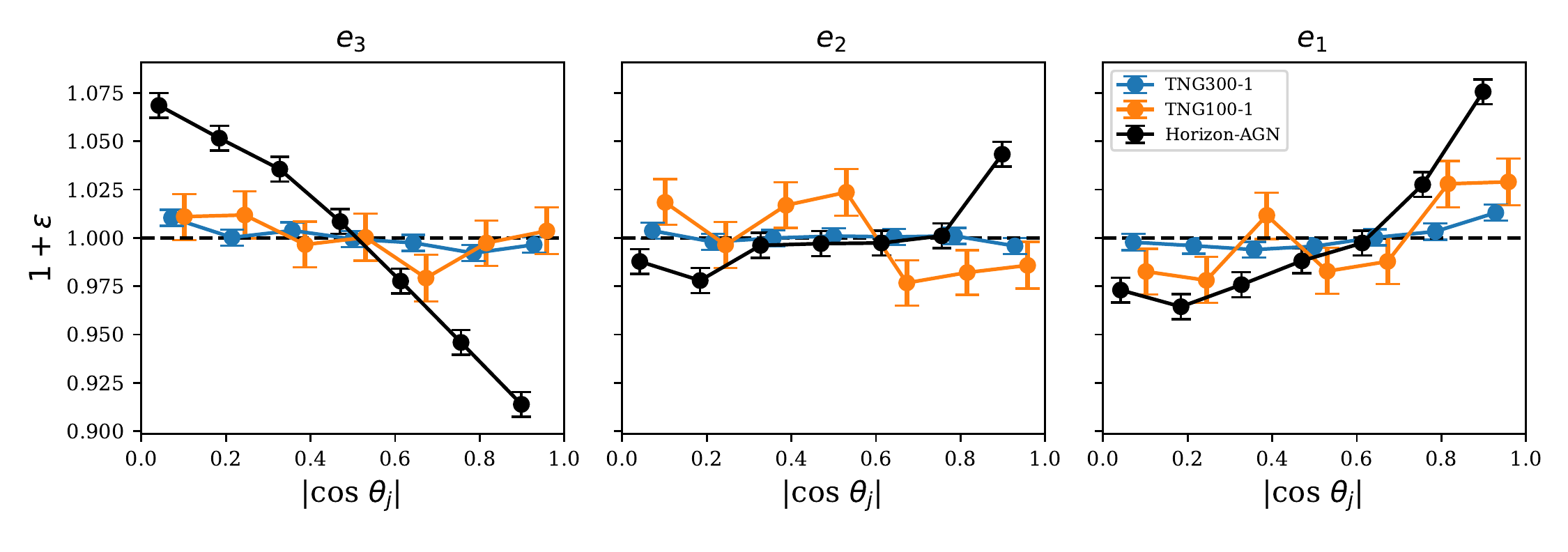}
                \caption{Comparison of Horizon-AGN ideal spin alignments from Figure 3 of \citet{Codis2015Spin} (in black) with corresponding spin alignments from TNG300-1 (blue) and TNG100-1 (red). TNG samples are derived from the samples described in Section \ref{sec:sim:methods:shapes}, with an additional minimum-mass cut of $M_* > 10^{8.06}$ \msunh{}. Errorbars for the TNG300-1/TNG100-1 alignments are derived from bootstrap sampling, while errors on \citet{Codis2015Spin} are Poisson, taken from the original figure.}
                \label{fig:Horizon-AGN-spin-comparison}
            \end{figure*}
            
            In Figure \ref{fig:Horizon-AGN-spin-comparison}, we compare TNG spin alignments to those found by \citet{Codis2015Spin} for the total (non-binned) Horizon-AGN galaxy sample, with corresponding errorbars (bootstrap for TNG alignments, Poisson for Horizon-AGN).
            Horizon-AGN displays a clear, high-amplitude negative trend for \ethree{} and positive trends for \etwo{} and \eone{} as expected from theory.
            Spin alignments with comparable amplitudes at \zone{} have also been observed for the SIMBA hydrodynamic simulation\footnote{However, note that the DISPERSE cosmic web finder \citep{DISPERSE} is used instead of the deformation tensor formalism for that work.} \citep{Kraljic20}.
            However, we find that alignment directions for TNG100-1 in the same mass range are more ambiguous, which can also be seen in Figure \ref{fig:z1-spin-ideal}.
            While TNG300-1 displays alignments in the same sign as theory and Horizon-AGN, their amplitudes are much lower than Horizon-AGN's.

            To quantify the level of disagreement between Horizon-AGN and TNG, we compute the ratio of \meanalign{} above null (0.5) of Horizon-AGN and TNG, $r_{\text{MA}}$:
            \begin{equation}
                r_{\text{MA}} = \frac{\meanalign{}_{\text{Horizon-AGN}} - 0.5}{\meanalign{}_{\text{TNG}} - 0.5}
            \end{equation}
            As we expect alignments for \ethree{} and \eone{} to be negatively/positively trending from theory for both Horizon-AGN and TNG, we only calculate $r_{\text{MA}}$ for alignments with those two eigenvectors.
            
            Between Horizon-AGN and TNG300-1, $r_{\text{MA}}$ for \ethree{} and \eone{} respectively is $(10.8, 6.9)$. Between Horizon-AGN and TNG100-1, $r_{\text{MA}} = (11.5, 2.4)$.
            Compared to Horizon-AGN's galaxy sample, TNG300-1 is missing some galaxies on the low mass end due to mass resolution limits.
            As we see in Figure \ref{fig:all-spin-ideal} (and expect from theory), lower-mass galaxies have less alignment strength, so we caution that our TNG300-1 $r_{\text{MA}}$ is likely overestimated.
            To be conservative, we therefore take the smallest ratio between \ethree{} and \eone{} for the TNG100-1/Horizon-AGN comparison to be the ``overall ratio of alignment strength" between Horizon-AGN and TNG, $r^*_{\text{MA}} = 2.4$.
            
            While a full analysis of the reasons behind this large alignment discrepancy is beyond the scope of this work, we speculate that this discrepancy may be partially caused by differences in feedback prescriptions between the simulations.
            \citet{TomassettiHaloGalMisalign} find that the alignment between galaxies and their surrounding halos has a large dependence on feedback strength.
            In particular, stronger feedback increases the strength of galaxy-halo alignment (and thus, galaxy-cosmic web alignment) at high redshifts, by delaying the time at which the baryonic component of galaxies compactifies.
            Observational measurements of high-redshift alignments could therefore be a very promising constraint on feedback models for simulations; and, in general, on their subgrid physics models.

\section{Mock Subaru PFS Surveys} \label{sec:mock}

    \subsection{Methods}
    \label{sec:mock:methods}
    
        In the previous section, we examined the level of alignment between the galaxies and underlying cosmic web within the IllustrisTNG simulations, which reflect the underlying `truth' in the simulations. 
        We next turn to the question of whether these alignments, e.g.\ as shown in Figures~\ref{fig:all-shape-ideal}, can actually be detectable in upcoming spectroscopic surveys.
        This is especially relevant in the context of the Galaxy Evolution component of the Subaru PFS Strategic Program survey \citep{pfs-whitepaper}, which will for the first time cover large enough cosmological volumes with sufficient fidelity to reconstruct the cosmic web at high redshifts. 
        We will construct mock galaxy observations and cosmic web reconstructions, adopting parameters planned for the PFS Galaxy Evolution Survey components at \zone{} and \ztwo{}.
        
        We first select a sample of galaxies that match the expected numbers of spectroscopic redshifts by matching their photometric magnitudes.
        We then displace the galaxies into redshift space, i.e.\ convolving their radial positions with their line-of-sight velocities  (Section \ref{sec:mock:methods:gal-sample}).
        For \zone{}, we infer the reconstructed dark matter density directly from the observed galaxy distribution (Section \ref{sec:mock:methods:z1-cweb}), while for \ztwo{}, we independently reconstruct the dark matter density using Lyman-alpha tomography (Section \ref{sec:mock:methods:z2-cweb}).
        The redshift-space distortion and both of these reconstruction methods depend on the viewing angle onto the simulation, which we marginalize over for reasons detailed in Section \ref{sec:mock:methods:proj}.
        For both redshifts, we calculate the deformation tensor (the reconstructed cosmic web) from the reconstructed dark matter density.
        
        Galaxy shapes for PFS galaxies can be estimated from deep high-resolution images at wavelengths covering their stellar components. Galaxy spins, the other hand, require resolved spectroscopy (typically with integral field unit spectrographs, or IFUs) to measure their rotations and hence angular momenta. 
        While it is in principle possible to obtain IFU data of galaxies at our $z\sim 1-2$ redshifts of interest, 
        the number of galaxies that would be need to be observed to cover a substantial fraction of the PFS sample needs to $>100\times$ larger than existing high-redshift IFU samples such as \citealt{KMOS3D_2019}. 
        We therefore do not expect the galaxy spin-cosmic web alignment at \zone{} and \ztwo{} to be measurable in the foreseeable future, and will focus solely on shape-cosmic web alignments in the remainder of this work.
        
        \subsubsection{Viewing Angle Variance}
        \label{sec:mock:methods:proj}
        
            When observing galaxies in reality, we are only able to observe their shape projected onto the 2D plane of the sky.
            While for disc-like galaxies it might be possible to estimate the inclination from the galaxy's detailed structure and therefore reconstruct 3D shape, this has only been attempted with reasonable effectiveness in the Local Universe \citep[e.g.,][]{HaynesSpinEst} --- we are certainly unable to resolve comparably detailed galaxy structure for large galaxy samples at $z\gtrsim 1$.
            In addition, this method is not applicable to elliptical-type galaxies, which typically have stronger alignment with the cosmic web \citep{Pandya}.
            For our mock observational survey analysis, we therefore project the shape of each galaxy onto the plane normal to a given viewing angle, which we approximate as the same across the simulation box\footnote{The angular scale subtended by a transverse comoving distance of $\sim 200\,\mpc$ (the extent of TNG300-1) at $z\sim 2$ is $\sim 3\deg$, for which small-angle approximations are still reasonable.}.
            
            For galaxy shapes, we start with the full galaxy reduced inertia tensor, which we interpret as an ellipsoid quadratic-form matrix.
            We project the inertia tensor ellipsoid onto the plane normal to the viewing angle, and take the new longest axis of the projected elliptical shadow as our projected shape vector.
            We also project all deformation tensor eigenvectors onto the viewing angle normal plane, and then calculate \alignment{}.
            As vectors with independent random directions in 3D reduce to vectors with independent random directions in 2D when projected onto a plane, we now use the alignment distribution of independently random 2D vectors as our null distribution.
            The distribution of $\cos\theta$ for such vector pairs is the beta distribution rescaled onto the interval $\cos\theta \in [-1, 1]$ (centered at 0) with distribution parameters $\alpha = \beta = 0.5$ [\citenum{BetaDistStackexchange}].
            The distribution of \alignment{} is the renormalized upper half of that beta distribution.
            
            However, in preliminary analysis we find that this projected alignment is significantly dependent on the selected viewing angle to the simulation volume. In other words, a given sample of galaxies in the simulation could vary significantly by choosing the viewing angle from which the 2D shape-cosmic web alignments are computed. When averaged over large numbers of viewing angles, the alignment signal varies significantly: in some cases, the mean alignment magnitude is reduced up to a factor of $\sim 2$ compared to the naive result computed from some of the individually strong viewing angles. We hereafter refer to this effect as ``viewing-angle variance''.

            \begin{figure*}
                \centering
                \includegraphics[width=0.7\textwidth]{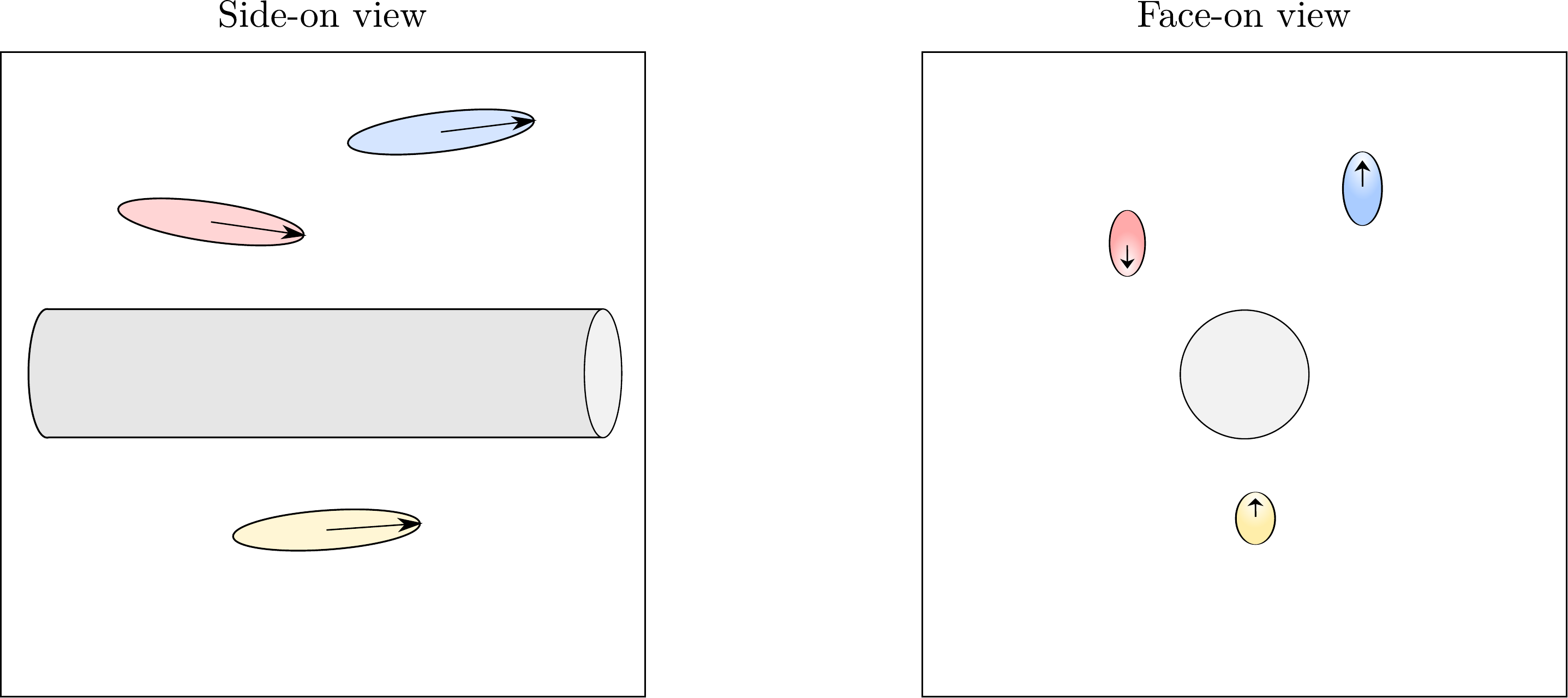}
                \caption{Cartoon illustrating viewing-angle variance on alignments caused by cosmic variance. For the small simulation volume depicted here with a filament running through it, galaxies are highly aligned with the filament in 3D. This extends to the projected alignment when looking ``side-on" to the filament (left panel). However, if we view the volume ``face-on" to the filament (right panel), the projected alignment is nearly random.}
                \label{fig:va-variance}
            \end{figure*}
            
            We hypothesize that this variation is a result of global anisotropies in the eigenvector orientations (i.e.\ dark matter flows) caused by cosmic variance within the limited simulation volumes.
            As a toy example, consider a small ($L \sim$ 3-10 \mpc{} side length) volume depicted in Figure \ref{fig:va-variance}, which is taken up by a single large filament passing through, with galaxies aligned with the filament.
            
            For viewing angles looking ``down the barrel" at the filament, projected eigenvectors would be sensitive to small perturbations in viewing angle, and vice versa for viewing angles looking length-wise at the filament\footnote{A similar effect is responsible for the systematics discussed in \citet{Lamman22}, although their application is very different from ours.}.
            This would therefore induce a high degree of viewing-angle variance in projected alignment and derived quantities such as mean projected alignment.
            Given this hypothesis, we expect the viewing-angle variance in mean projected alignment to scale inversely with volume size.
            
            We investigate the possibility of this inverse scaling relationship between volume size and viewing-angle variance using the ``true" IllustrisTNG data discussed in Section \ref{sec:sim:methods}, for \zone{}.
            To do so, we project the galaxy shape vectors and deformation tensor eigenvectors onto 64 evenly-spaced viewing angles into the volume.
            Next, we divide the $L=205 \,\mpc{}$ side-length volume into $N=$\{1, $2^3$, $4^3$, $8^3$\} evenly-sized, non-overlapping sub-cubes.
            For each sub-cube, we consider the $64 \times N$ projected alignments computed from the $N$ galaxy-eigenvector pairs within the sub-cube.
            For \textit{each viewing angle individually}, we compute the mean projected alignment across all galaxy-eigenvector pairs.
            Finally, we compute the standard deviation of those 64 mean projected alignments, which we take as a measure of the viewing-angle variance for that sub-cube, $\sigma_{VA}$.

            \begin{figure}
                \centering
                \includegraphics[width=0.5\textwidth]{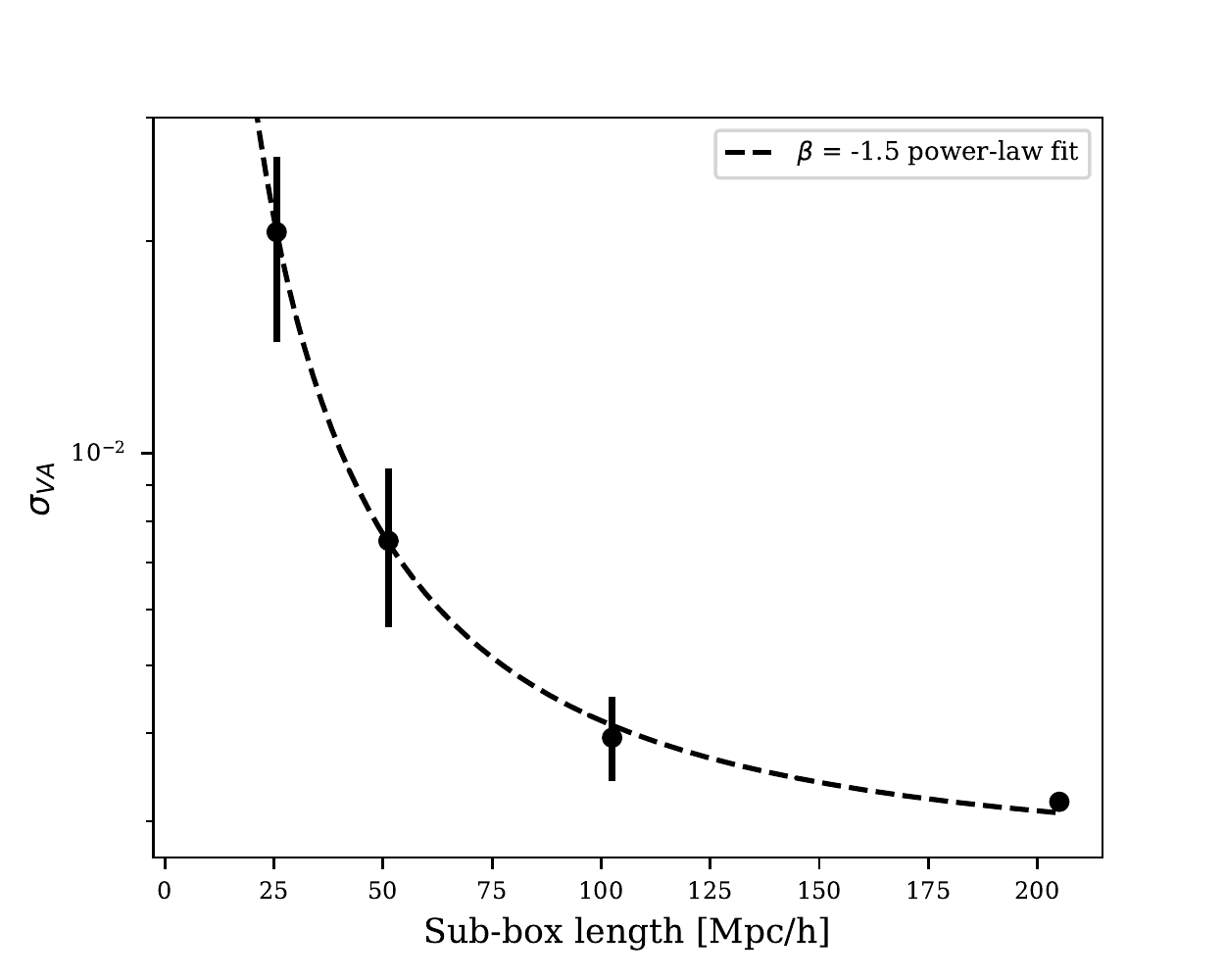}
                \caption{Standard deviation of projected \zone{} real-space mean alignment $\sigma_{VA}$ computed from 64 viewing angles of the same sub-volumes, plotted as a function of subvolume side length. As we divide the TNG300 volume into non-overlapping subvolumes, we can also plot the error on $\sigma_{VA}$. The data are consistent with a power-law with fixed exponent $\beta = -1.5$, which is equivalent to a $\beta = -0.5$ power-law against volume.}
                \label{fig:viewing-angle-variance}
            \end{figure}
            
            Figure \ref{fig:viewing-angle-variance} displays the results of this analysis.
            Since for box sizes besides $L=205^3$ $\mpccubed{}$ we have multiple sub-cubes, we display the mean and standard deviation across sub-cubes of $\sigma_{VA}$.
            To avoid using a logarithmic scale on the x-axis, we plot $\sigma_{VA}$ against sub-cube side length instead of sub-cube volume.
            As expected, $\sigma_{VA}$ increases with decreasing side length/decreasing volume.
            To determine the exact relationship, we try fitting a power-law with fixed exponent $\beta = -1.5$ to the mean $\sigma_{VA}$ (rather than the individual $\sigma_{VA}$ data points to avoid overweighting small volumes).
            We find the data are consistent with this $\beta = -1.5$ power-law against side length, implying a $\beta = -0.5$ power-law against volume.

            The strength of this viewing angle variance even in the TNG300 volume suggests that even $L=205\,\mpc$ 
            simulation box sizes are affected by cosmic variance of large-scale filaments as far as galaxy-cosmic web alignments are concerned. In our mock survey analysis, we therefore marginalize over viewing angles by adopting a viewing angle-agnostic approach.
            Specifically, we generate 64 evenly spaced points on the unit half-sphere using a Fibonacci lattice \citep{Hannay_2004}.
            We take the unit vectors connecting those points and the origin as our set of observer viewing angles into the TNG300-1 simulation.
            For every viewing angle, we construct mock galaxy observations with the procedure detailed in the rest of the methods section.
                
        \subsubsection{Galaxy sample selection}
        \label{sec:mock:methods:gal-sample}
            
            For our mock observational survey analysis only, we subselect the galaxies in the simulation to match the expected sample from the Subaru Prime Focus Spectrograph Galaxy Evolution (hereafter PFS GE) survey, which is planned to take place over $\sim 100$ nights of observation on the Subaru Telescope.
            At $z = 0.7 \sim 1.7$, the possible targets for the PFS GE Survey number $N_g \sim$ 357,000 galaxies over a total survey volume of $V_S \sim$ \num{32467532.467532467} \mpccubed{}, and at $z = 2.1 \sim 2.5$, $N_g \sim$ 100,000 galaxies over a volume of $V_S \sim$ \num{2.7e7} \mpccubed{} \citep{pfs-whitepaper}.
            
            As only some rest-frame galaxy magnitudes and not full synthetic galaxy spectra are provided in the IllustrisTNG release, we proceed to define the equivalent targets in the simulated IllustrisTNG catalog by performing an abundance-matching based on their number densities and ranked-ordered luminosities within a given filter wavelength range. 
            For the \zone{} redshift range, the PFS survey target sample are selected based on Hyper Suprime-Cam $Y$-band limiting magnitude (centered at approximately $\lambda_C \approx $ \SI{1}{\um} wavelength),
             so we use the rest-frame Johnson $V$-band ($ \lambda_C \approx 0.551$ \SI{}{\um}) photometric magnitude supplied in the IllustrisTNG catalog for each galaxy as an approximate proxy for the observed-frame $Y$-band magnitude.
            This approach ignores filter bandpass shapes, and so should be considered approximate; additionally, the intrinsic magnitudes given by IllustrisTNG do not include the effects of dust extinction.
            Using the V-band magnitude, we select the brightest 94,766 galaxies in our original sample, to match the expected number density of possible PFS $z\sim 1$ survey targets: $n_g = 0.011 \invmpccubed{}$.
            
            We perform the same procedure for selecting survey galaxies at the \ztwo{} range, which are part of the `IGM foreground' galaxy sample designed to provide a sample of galaxies within the volume probed by IGM tomography from background galaxy and quasar Lyman-$\alpha$ forest sightlines. 
            Here, observed-frame $J$-band at $\lambda_C \approx 1.22$ \SI{}{\um} is the limiting magnitude in target selection, which we approximate by selecting the 31,908 brightest galaxies in rest-frame $B$-band magnitude ($\lambda_C \approx 445$ nm)
            We match the selection of the targets to the number density of candidates, which is $n_g = 3.7 \times 10^{-3} \, \invmpccubed{}$.

            We emphasize that the number of possible survey targets does not equal the number of galaxies which will have successful redshift measurements.
            PFS expects $\sim$70\% of observed targets to yield successful redshifts, for a final $z\sim 1$ data sample of $N_g \approx$ 250,000 galaxies.
            For $z\sim 2$, 22\% of possible targets will actually be observed (i.e.\ the sampling rate), and of those targeted, 70\% are expected to yield successful redshifts.
            This translates to a final data sample of $N_g \approx $ 15,400 galaxies at $z\sim 2$.
            We also consider a scenario where the \ztwo{} component is allocated double the baseline observing time within the survey, which increases the sampling rate to 44\%, and the final data sample to 30,800 galaxies.
            Although the spectroscopic redshift success rate will depend on a variety of factors in the real observations, here we will make the simplifying assumption that every galaxy above the magnitude limit has an equal chance to yield a successful redshift for our analysis.
            Due to this assumption, it is valid for us to sample without replacement the \textit{absolute} number of successful galaxies PFS expects from our simulation's possible target pool, even if the number density of galaxies we select is higher than the actual number density of successful galaxies that PFS will observe.
            We therefore create two samples with 15,400 and 30,800 galaxies for \ztwo{} out of the 31,908 possible targets (ensuring that the former sample is a strict subset of the latter).
            For the remainder of this work, we refer to these two samples as \ztwonormal{} (``\ztwo{} fiducial") and \ztwodouble{} (``\ztwo{} with 2$\times$ greater number density") respectively.
            The $N_g =$ 250,000 observed galaxies expected at \zone{} in fact exceeds the total number of galaxies within TNG300-1 that have the correct luminosities. Therefore we use all 94,766 possible targets within the TNG300-1 volume and perform a post-hoc rescaling of the uncertainties. This will be described in Section \ref{sec:mock:methods:errs}.
            For the remainder of the work, we refer to this $z\sim 1$ dataset as \zonenormal{}.

        \subsubsection{\zone{} cosmic web reconstruction}
        \label{sec:mock:methods:z1-cweb}
        
            At \zone{}, the expected number density of spectroscopically-confirmed galaxies from PFS is high enough to enable accurate reconstruction of the underlying 3D matter density field using the observed galaxy positions as a biased tracer of density, as in redshift surveys of the Local Universe ($z\lesssim 0.1$, \citet{kitaura+09,elucid,BORG}).
            For density reconstruction, we use the \tardis{} density reconstruction code \citep{TARDISI,TARDISII}, which uses maximum-likelihood analysis in conjunction with a gravitationally evolved forward model of the matter density field.
            \tardis{} constrains its search space to density configurations that could have evolved from initial conditions, and has been shown to accurately reconstruct the cosmic web eigenvectors of the deformation tensor.
            We use \tardis{}'s ``galaxy-only" density reconstruction mode, in which the matter density is computed from the distribution of observed galaxies in the forward model using fixed bias parameters in Lagrangian space.
            In order to fit these bias parameters, we match PFS's expected number density by uniformly subsampling without replacement 70\% of our \zone{} magnitude-limited data sample, which is 66,336 galaxies.
            With this galaxy subsample, we estimate the bias parameters to be $b_1 = 1.8$, $b_2 = 0.1$.
            
            For each viewing angle, we take the same galaxy subsample used to fit the biases, compute the RSD along the viewing angle's line of sight, and input into \tardis{}.
            We opt not to select different subsamples for each viewing angle because the purpose of varying the viewing angle is to quantify the effect of viewing angle alone on the alignment.
            For all viewing angles, we reconstruct the density on a 384-cell length cube, with box length of $L=$205 \mpc{}.
            We then smooth the reconstructed real-space density of \tardis{} with a 2 \mpc{} Gaussian kernel, and compute the deformation tensor for each smoothed, reconstructed volume.
        
        \subsubsection{\ztwo{} cosmic web reconstruction}
        \label{sec:mock:methods:z2-cweb}
        
            In order to reconstruct the cosmic web at \ztwo{} where only a relatively low number density of galaxies will be observed, we carry out a joint reconstruction with Lyman-$\alpha$ forest tomography (also known as IGM tomography) as the primary tracer of the cosmic web, with galaxies included as a secondary tracer.
            In Ly$\alpha$ forest tomography, redshifted hydrogen Ly$\alpha$ ``forest" absorption lines in the spectra of background sources are used to infer the neutral density of diffuse IGM hydrogen along the line of sight.
            As first argued by \citet{LeeLyaPOC}, with a sufficient transverse density of background quasars and UV-bright star-forming galaxies, the 3D IGM density field can be reconstructed to a 2-3 \mpc{} spatial resolution.
            The COSMOS Lyman-alpha Mapping and Tomography Observations (CLAMATO, \citealt{CLAMATO_DR1}) survey has to-date performed Ly$\alpha$ tomographic reconstruction over a 0.21 deg$^2$ footprint within the COSMOS field, reconstructing the IGM density at a spatial resolution of 2.5 \mpc{} over a survey volume of $V_S = $\num{4.1e5} \mpccubed{} \citep{CLAMATO_DR2}.
            The sources used had an average on-sky transverse separation of 2.4 \mpc{} at \ztwo{}.
            PFS aims to perform Ly$\alpha$ tomographic reconstruction using a similar background source separation, but for a much larger total volume of $V_S \sim$ \num{2.7e7} \mpccubed{} (an ongoing intermediate-sized survey is the LATIS survey, \citet{LATIS}).
            While the median redshift of the PFS IGM tomography map is expected to be around $\langle z \rangle \approx 2.5$, this represents only around a 700 Myr difference in cosmic time compared to $z=2$, so we expect the overall trends we find in the IllustrisTNG $z=2$ snapshot to qualitatively reflect the PFS survey if not quantitatively exact.
            
            For each viewing angle, we construct a ``joint" galaxy and Ly$\alpha$ tomographic survey using mock data derived from the TNG300-1 simulation.
            For the galaxy portion, we again match expected galaxy number densities through random subsampling of our abundance-matched, magnitude-limited galaxies.

            The Ly$\alpha$ tomographic portion of our mock survey is composed of a number of ``skewers" that record relative Ly$\alpha$ absorption in a background object's spectrum, along the viewing angle's line of sight.
            Since IllustrisTNG is a hydrodynamic simulation, we directly construct mock Ly$\alpha$ absorption skewers using the \fakespec{} code \citep{BirdFakeSpectra}. %

            We follow expected PFS survey parameters by generating skewers at random sky positions with a number density such that the average transverse separation between skewers is 2.4 \mpc{}. These skewers do not initially contain instrumental noise.
            Skewers which contain a damped Ly$\alpha$ absorber (DLA), i.e.\ with integrated HI column density of $N_\mathrm{HI} \geq $ \SI{2e20}{\cm^{-2}} are filtered out since we expect to be able to remove them in observational data.
            We then rebin the skewers to an instrumental spectral resolution of 0.7 \SI{}{\angstrom} per pixel, and simulate spectral dispersion by smoothing the skewers with a 1D Gaussian kernel of size 2.7 \SI{}{\angstrom}.
            This matches the expected instrumental performance of the PFS spectrograph's blue arm, which at a range of 380-650 nm covers the observed Ly$\alpha$ spectra \citep{pfs-instrument}.
            Next, we add instrumental noise to the skewers using the noise model described in \cite{TARDISI,TARDISII} %
            and add an offset to the overall transmitted flux in each skewer to model systematic fitting errors of the skewer's background source continuum \cite{krolewski:2017}.
            
            To reconstruct the underlying dark matter density from our mock skewers, we apply \tardis{}  jointly to the galaxy and Ly$\alpha$ tomography data \citep{TARDISII} with the likelihood now calculated from a weighted combination of the galaxy density tracer and the Ly$\alpha$ absorption tracer.
            For all viewing angles, we again reconstruct the density on a 384-cell length cube and smooth with a 2 \mpc{} Gaussian kernel before computing the deformation tensor.

        \subsubsection{Error modeling}
        \label{sec:mock:methods:errs}
        
            In the mock PFS survey, our galaxy redshift-space positions and projected shapes depend on the viewing angle.
            Both \zonenormal{} and \ztwonormal{} density reconstructions also depend on the viewing angle: RSD is applied to the input galaxies along the viewing angle line of sight, while for \ztwonormal{}, Lyman-alpha skewers are also created along each line of sight.
            In order to quantify the effects of differing viewing angles, when we compute alignment statistics for a sample or subsample of $N_g$ galaxies, we marginalize over all viewing angles.
            For a given eigenvector, we therefore compute alignment statistics on 64$\times N$ galaxy-nearest eigenvector pairs.
            
            Since the actual survey will only be on (approximately) one viewing angle but over a larger overall volume than that of TNG300, we adopt a modified bootstrap sampling scheme to quantify the viewing angle variance in addition to our usual estimate of cosmic variance.
            For each bootstrap iteration, we perform the following procedure.
            First, we randomly select one viewing angle, and consider only the galaxy-eigenvector pairs which use that viewing angle, for RSD, density reconstruction, and shape projection (if applicable).
            We then perform normal bootstrap sampling on the pairs, and compute a histogram on \alignment{} along with \meanalign{}.
            The 16th and 84th percentiles on the computed histogram bins and \meanalign{} from 10,000 bootstrap sampling iterations (with a random viewing angle for each iteration) are then adopted as our error bars.
            
            As mentioned in Section \ref{sec:mock:methods:gal-sample}, our pool of 94,766 possible survey targets within the TNG-300-1 volume is not large enough to sample the $N_g$ = 250,000 galaxies expected for the \zone{} survey planned for Subaru PFS GE.
            We therefore perform bootstrap error estimation on all survey targets, and scale the resulting errorbars by $\sqrt{94,766 / 250,000} \simeq 0.615$.
            This is an approximate scaling assumption based on the Central Limit Theorem's expression for sample standard deviation, which scales as 1/$\sqrt{N}$ with a sample size $N$.
            
            In Section \ref{sec:mock:methods:proj}, we found that the variance in projected alignment across differing viewing angles scales as $\sigma_{VA} \propto V_S^{-1/2}$, where $V_S$ is the survey volume.
            We therefore additionally scale the errorbars of the projected alignment only by $\sqrt{205^3 \mpccubed{} / V_{S}}$, where $V_{S}$ is the expected volume of the survey.
            For \zone{}, $V_S \sim \num{32467532.467532467}$ $\mpccubed{}$, and $V_S \sim \num{2.7e7}$ $\mpccubed{}$ for \ztwo{}.
            This yields an error scaling factor of 0.515 and 0.565 respectively for \zone{} and \ztwo{}.

    \subsection{Results and Discussion} \label{sec:mock:results}

    The first step of validating the results is to test the recovery of the cosmic web from the density reconstructions described in the previous subsections. Since this largely repeats similar exercises previously done in \citet{KrowleskiSpin} and \citet{TARDISII}, we move this discussion to Appendix \ref{sec:appendix:recon}. In all cases, the cosmic web eigenvectors derived from the reconstructed density fields are in excellent agreement with the underlying `truth', which sets us up for the subsequent analysis.

    \begin{table*}[ht!]
    \centering
    \begin{tabular}{lllll}
    \hline
    Sample      & Mass threshold           & $N_{gal}$ & 3D $\sqrt{\chi^2}$                                                                                  & Projected $\sqrt{\chi^2}$                                                                           \\ \hline
    \zonenormal{}           & N/A                      & 250000*   & \num[scientific-notation=false,round-mode=figures,round-precision=\mainsigfigs]{5.42682}            & \num[scientific-notation=false,round-mode=figures,round-precision=\mainsigfigs]{5.321684971945624}  \\
                  & $\leq 10^{10.1} \msunh{}$ & 122056*   & \num[scientific-notation=false,round-mode=figures,round-precision=\mainsigfigs]{1.17019}            & \num[scientific-notation=false,round-mode=figures,round-precision=\mainsigfigs]{3.491704810046807}  \\
                  & $> 10^{10.1} \msunh{}$    & 127944*   & \num[scientific-notation=false,round-mode=figures,round-precision=\mainsigfigs]{6.80559}            & \num[scientific-notation=false,round-mode=figures,round-precision=\mainsigfigs]{4.520962629900436}  \\ \hline
    \ztwonormal{} & N/A                      & 15400     & \num[scientific-notation=false,round-mode=figures,round-precision=\mainsigfigs]{0.9446235084441081} & \num[scientific-notation=false,round-mode=figures,round-precision=\mainsigfigs]{1.3047330881408605} \\
                  & $\leq 10^{10.5} \msunh{}$ & 11264     & \num[scientific-notation=false,round-mode=figures,round-precision=\mainsigfigs]{0.3723374043683161} & \num[scientific-notation=false,round-mode=figures,round-precision=\mainsigfigs]{0.8841836499054957} \\
                  & $> 10^{10.5} \msunh{}$    & 4136      & \num[scientific-notation=false,round-mode=figures,round-precision=\mainsigfigs]{1.056814897993778}  & \num[scientific-notation=false,round-mode=figures,round-precision=\mainsigfigs]{1.1883086509888316} \\ \hline
    \ztwodouble{} & N/A                      & 30800     & \num[scientific-notation=false,round-mode=figures,round-precision=\mainsigfigs]{1.4980471911473852} & \num[scientific-notation=false,round-mode=figures,round-precision=\mainsigfigs]{1.471300765505908}  \\
                  & $\leq 10^{10.5} \msunh{}$ & 22652     & \num[scientific-notation=false,round-mode=figures,round-precision=\mainsigfigs]{0.7605272976896116} & \num[scientific-notation=false,round-mode=figures,round-precision=\mainsigfigs]{0.9469380458321802} \\
                  & $> 10^{10.5} \msunh{}$    & 8148      & \num[scientific-notation=false,round-mode=figures,round-precision=\mainsigfigs]{1.5748194553526143} & \num[scientific-notation=false,round-mode=figures,round-precision=\mainsigfigs]{1.4284871440963192} \\ \hline
    \end{tabular}
    \caption{3D and projected significances ($\sqrt{\chi^2}$) for alignments between galaxy shapes and cosmic web eigenvectors calculated from the mock observational surveys, at \zone{} and \ztwo{}. We note that $N_{gal}$ is the number of galaxies that PFS is predicted to observe for each redshift. For \zone{}, the stated $N_{gal}$ (marked with asterisks) is $\sim$2.6x the number of galaxies actually used in our analysis, due to volume limitations on TNG300-1. We therefore apply additional error scaling to compensate, as detailed in Section \ref{sec:mock:methods:errs}.}
    \label{tab:mock-sig-excerpt}
    \end{table*}
            
        \subsubsection{\zone{} mock observational survey alignment}
            
\begin{figure*}[ht]
                \centering
                \includegraphics[width=\textwidth]{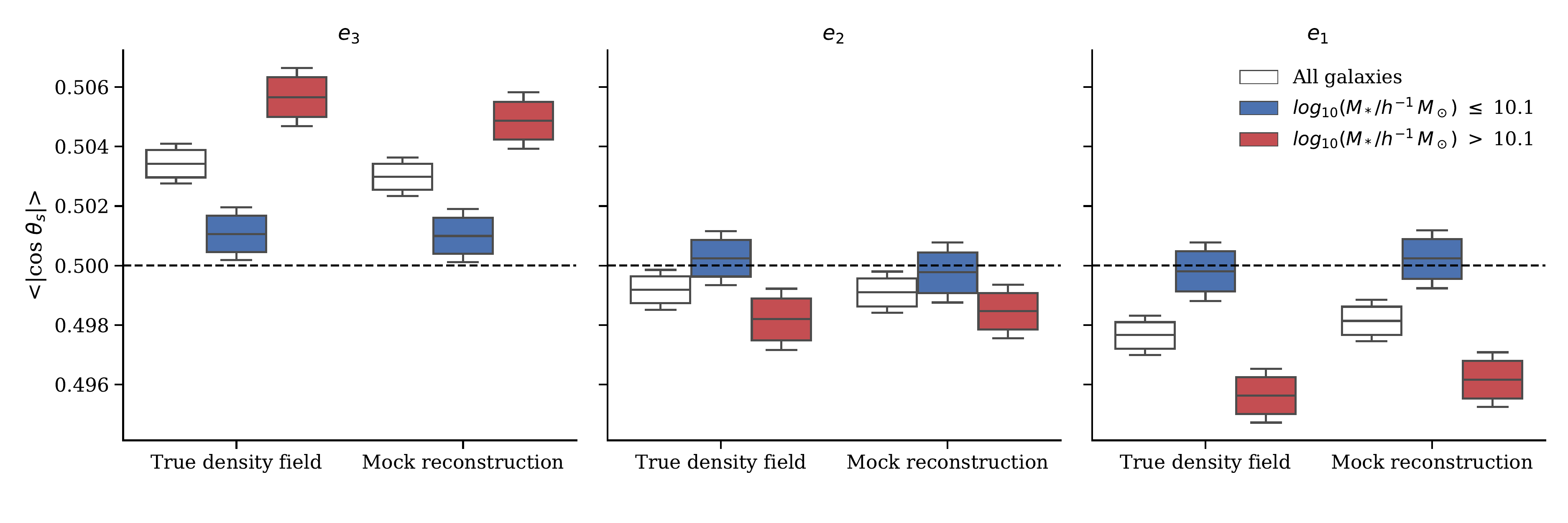}
                \includegraphics[width=\textwidth]{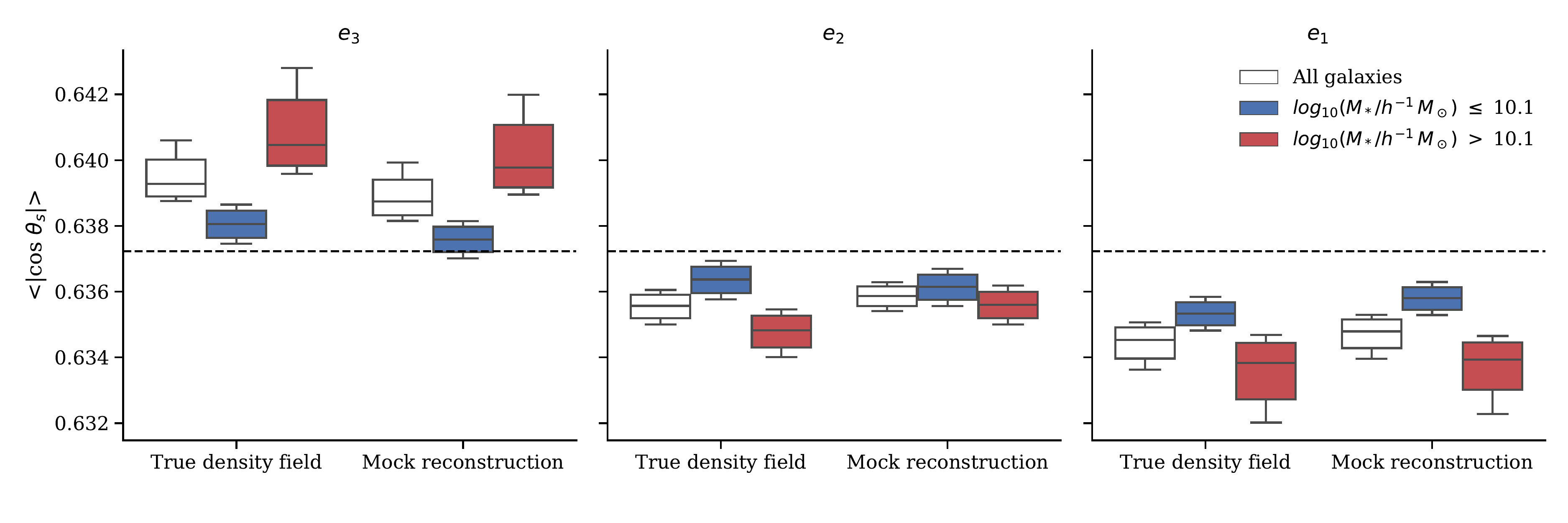}
                \caption{\zonenormal{} \maplotcap{}}
                \label{fig:z1-ma}
            \end{figure*}

            In Figure \ref{fig:z1-ma}, we present results for our mock survey alignment analysis for the \zonenormal{} galaxy sample at \zone{}. 
            In order to isolate the various observational effects on any potential measurement, we measure the alignment relative to both the `true' cosmic web, as well as that derived from the reconstructed mock density field. 
            
            In the left half of each subpanel, we show the mean alignment, \meanalign{}, of the galaxies relative to the cosmic web eigenvectors computed from the true density field. %
            The right half of each subpanel shows \meanalign{} computed from eigenvectors in the \tardis{} reconstructed density fields, that were applied to realistic mock data as detailed in Section \ref{sec:mock:methods:z1-cweb}.
            We hereafter refer to this as the ``mock reconstruction" alignment.
            In both cases, all measurements are calculated in redshift space.
            
            The top row of panels in the Figure~\ref{fig:z1-ma} is the \meanalign{} computed for the 3D galaxy shapes, while the bottom row is the \meanalign{} from the projected (2D) galaxy shapes and eigenvectors calculated over different viewing angle planes. In every subpanel, a dashed black line represents the \meanalign{} from the null distribution.
            As described in Section \ref{sec:mock:methods:errs}, we marginalize over all viewing angles to derive errors for \meanalign{}, and additionally scale our errors to correct for galaxy number counts (for both 3D and projected alignment) and for viewing-angle variance (for projected alignment only).
            For the alignments computed from the true density field, the differing redshift-space galaxy displacements along different viewing angles are accounted for in the bootstrap error estimation process; and for the mock reconstruction alignment, the differing density reconstructions are accounted for as well.
            In addition to presenting \meanalign{} for the full samples, we also calculate \meanalign{} with the sample split into low and high stellar mass samples.
            We set the threshold mass to the transition masses found in Section~\ref{sec:sim:results:shape}, i.e. $M_*=[3.16\times 10^{10},\; 1.26\times 10^{10}]\,\msunh$ at $z=[1,2]$ respectively.
            For the purpose of assessing observational feasibility, we use the stellar mass as a rough proxy for brightness.
            While TNG includes synthetic magnitudes for galaxies, we opt not to split our sample by magnitude, as we are unable to incorporate the effects of dust extinction in our viewing-angle agnostic approach.

            In order to quantify the overall detection significance relative to the null distribution, we adopt a metric based on the reduced chi-squared statistic.
            We calculate the covariance of \meanalign{} across all three eigenvectors using bootstrap sampling, and define our significance metric as:
            \begin{equation}
            \label{eq:sig}
                \sqrt{\chi^2} = \sqrt{(\vec{m} - \vec{n})^T C^{-1} (\vec{m} - \vec{n})}
            \end{equation}
            where $\vec{m}$ is the vector of \meanalign{} with the elements populated by the values from the three eigenvectors, $\vec{n}$ is a constant vector of the null distribution's mean alignment, and $C$ is the covariance matrix.
            For our analysis of \zonenormal{}, the covariance matrix is computed using the rescaled 16th and 84th percentile bootstrap errors computed for the mean alignments.
            We present mock reconstruction alignment significances in Table \ref{tab:mock-sig-excerpt}.

            In 3D, the alignments seen between the galaxy shapes and cosmic web eigenvectors still preserve the underlying trends seen in Section \ref{sec:sim:results}.
            When shape alignments are split by the transition mass found in Section \ref{sec:sim:results:shape}, the low-mass sample has near-null alignment as expected, while the high-mass alignment is stronger than the total alignment.
            As \tardis{} yields a high fidelity reconstruction of the cosmic web eigenvectors in the volume, the mock reconstruction alignment is nearly identical to true density field alignment, with the only notable difference being a negligible ($\Delta$\meanalign{}$\sim$0.001) decrease in mean alignments.
            
            When the 3D alignments are projected into 2D, the general alignment trends as a function of stellar mass remain the same as the 3D case.
            Projection tends to increase the scatter of \meanalign{} due to the viewing-angle variance effect described in Section \ref{sec:mock:methods:proj}, but the volume rescaling we apply to the projected \meanalign{} appears to largely counteract the variation increase. 
            We note that the viewing angle projection also appears to introduce an error asymmetry into the mean alignments for both \ethree{} and \eone{}, with increased variation in the sign-direction of alignment (positive for \ethree{} and negative for \eone{}).
            This may be caused by a bias in eigenvector direction in the IllustrisTNG volume used, arising from cosmic variance as explored in Section \ref{sec:mock:methods:proj}.
            
            The effect of projection on the mock reconstruction alignment can be quantified in terms of the significance metric \sig{} listed in Table \ref{tab:mock-sig-excerpt}.
            Due to the volume error scaling, \sig{} is slightly decreased by a factor of $\sim$0.66 for the high-mass samples, and remains almost constant when considering the total (low-mass and high-mass) sample. %
            The 5.3$\sigma$ significance found in the galaxy shape alignments with the reconstructed cosmic web in the \zonenormal{} sample, when combined over all stellar masses, crosses the conventional detection significance threshold of 4$\sigma$, so we anticipate a successful detection in the PFS survey. %

        \subsubsection{\ztwo{} mock observational survey alignment}

            \begin{figure*}[ht]
                \centering
                \includegraphics[width=\textwidth]{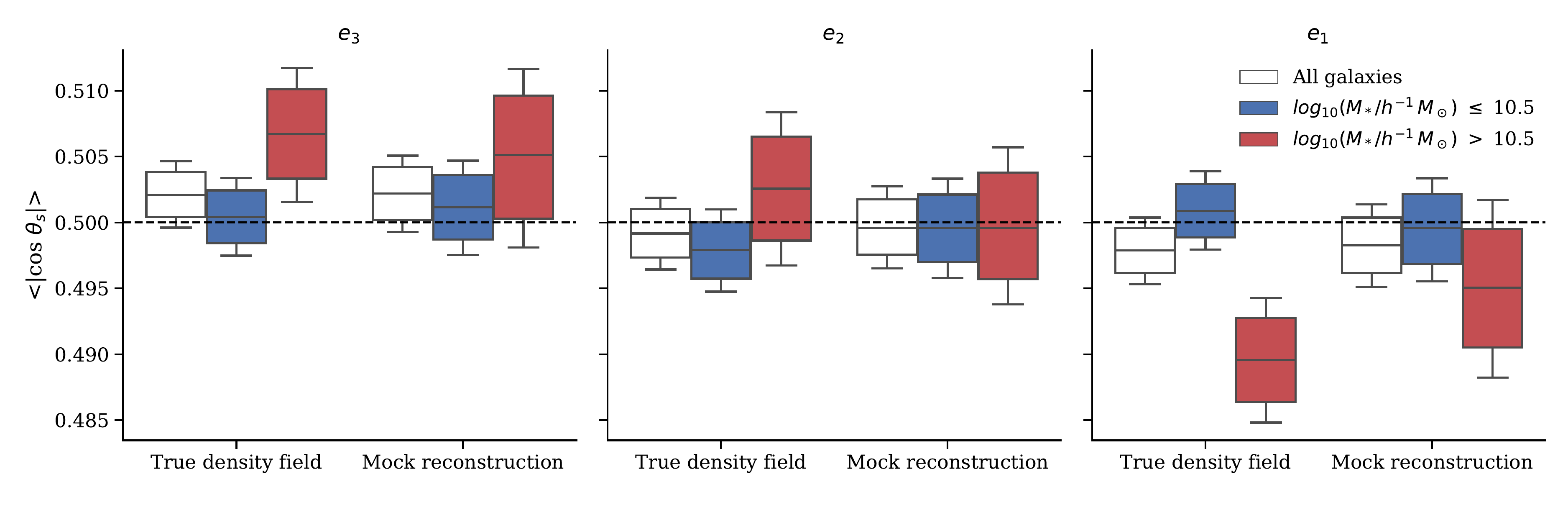}
                \includegraphics[width=\textwidth]{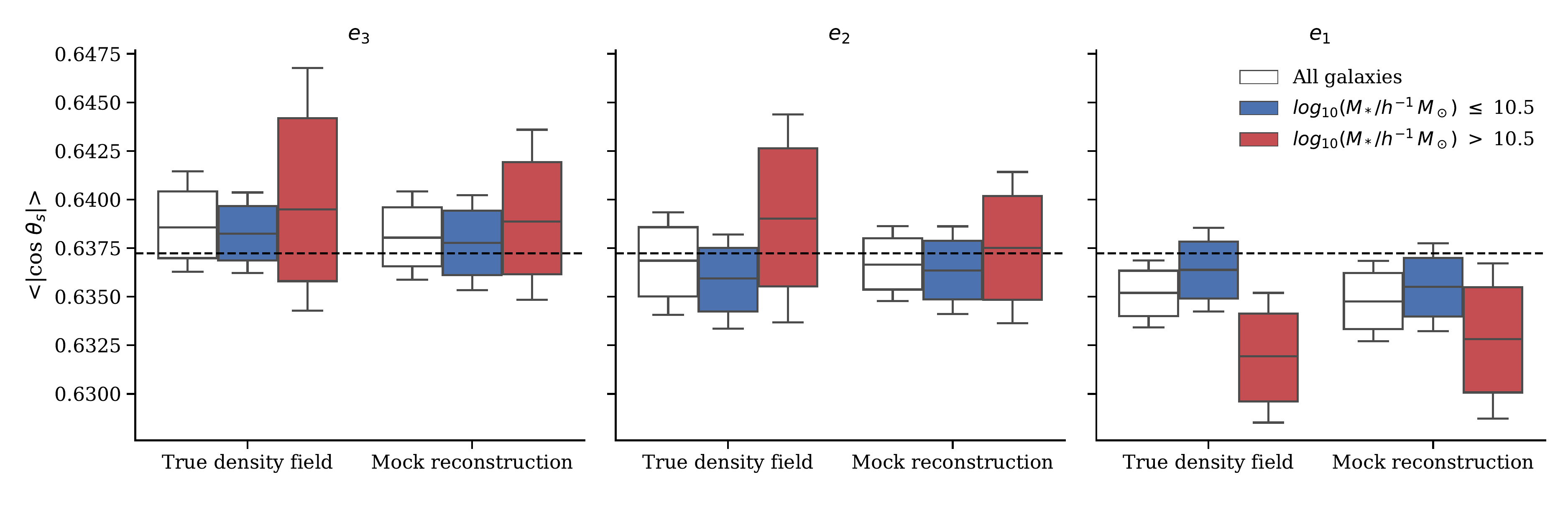}
                \caption{\ztwonormal{} \maplotcap{}}
                \label{fig:z2-ma}
            \end{figure*}
            
            At \ztwo{}, we display the mean alignment results for only \ztwonormal{} for visual clarity, though we will later discuss mock reconstruction results for the increased sample size (\ztwodouble{}).
            In Figure \ref{fig:z2-ma}, we show shape alignments for \ztwonormal{}.
            We find that even in 3D, the relatively small galaxy sample size produces significant uncertainties in both samples compared to the \zonenormal{} alignments.
            While the \meanalign{} central values for \ethree{} and \eone{} are comparable to those for \zonenormal{}, the central 68th percentile spread increases from $\Delta$\meanalign{} $\sim$ 0.002 to $\Delta$\meanalign{} $\sim$ 0.005.
            This increase is also greater for the high-mass sample, as the proportion of high-mass galaxies is $\sim$ 27\% at \ztwo{}, as opposed to $\sim$ 50\% for \zone{}.
            Projection further compounds this alignment degradation, with central \meanalign{} dropping below 1$\sigma$ for \ethree{} and \eone{} in all cases.
            As opposed to \zonenormal{} where high reconstruction quality has no visible effect on 3D alignment spread, the volume error scaling is unable to compensate for the comparatively poorer reconstruction quality in \ztwonormal{}.
            We therefore see a small increase in the central 68th percentile spread of $\Delta \meanalign{} \sim$ 0.004 for the mock reconstruction alignment, compared to the true density field alignment.
            In fact, the volume error scaling we apply may be detrimentally amplifying variance in the mean alignment caused by the small sample.
            When considering the mock reconstruction significance shown in Table \ref{tab:mock-sig-excerpt}, we find that the projected significance is unexpectedly \textit{increased} in comparison to the 3D significance for \ztwonormal{}.
            However, \ztwodouble{} shows an equal significance for projected and 3D mock reconstruction alignment, lending credence to the small-sample amplification effect.
            For both sample sizes and all mass splits, the mock reconstruction shape alignment significance is $<$1.5, so we do not anticipate a significant detection of the IllustrisTNG shape-cosmic web alignment at \ztwo{} by PFS.

        \subsubsection{Horizon-AGN Observability Estimate}

            Due to the large difference in spin-cosmic web alignment strength between IllustrisTNG and Horizon-AGN which we pointed out in Section \ref{sec:sim:results:horizon}, in this section we aim to provide a rough estimate on the detectability of the Horizon-AGN alignment signal.
            
            As the Horizon-AGN mean alignment signal is only published for \zone{} spin-cosmic alignment by \citet{Codis2015Spin}, we first take the ratio between Horizon-AGN and IllustrisTNG signal strength at \zone{} $r_{\text{MA}}$ that we have found in Section \ref{sec:sim:results:horizon}.
            This is $r^*_{\text{MA}} = 2.4$.
            We find in Appendix \ref{sec:appendix:shape-spin} that spin alignments and shape alignments in IllustrisTNG are highly correlated, as expected.
            We therefore use the spin-alignment signal ratio as a very rough guide for the expected detection significance of the Horizon-AGN \emph{shape}-cosmic web alignment signal at both \zone{} and \ztwo{}, by multiplying the IllustrisTNG shape ``\meanalign{} over null" $\vec{m} - \vec{n}$ by the same ratio.
            As $\sig{} \propto |\vec{m} - \vec{n}|$, this is equivalent to multiplying \sig{} by that factor.
            This assumes that the significance covariance matrix (see Equation \ref{eq:sig}) that we have calculated for IllustrisTNG alignments remains exactly the same if using Horizon-AGN.
            We believe this to be an appropriate conservative assumption, as we find in Figure \ref{fig:Horizon-AGN-spin-comparison} that the error bars for Horizon-AGN are much smaller than for the corresponding IllustrisTNG sample (though note that Horizon-AGN uses Poisson errors, while we use bootstrap errors for TNG).
            
            Following this procedure, we would expect the Horizon-AGN alignment signal to be detectable at $\sim3\sigma$ at both redshift bins, with the lowest total significance of \ztwonormal{} improving from $1.3\sigma$ to $3.1\sigma$ and the \zonenormal{} total significance increasing from $5.3\sigma$ to $12.7\sigma$.
            However, we note again that this is a very rough estimate, but does suggest that the Subaru PFS Galaxy Evolution Survey would be able to strongly detect or rule out the level of galaxy-cosmic web alignment predicted by Horizon-AGN.

        \subsubsection{Possible survey improvements}
            
            Assuming that detection significance scales as $\propto\sqrt{N_{gal}}$, we estimate that in order to constrain \ztwo{} IllustrisTNG shape alignments with a significance of $\sig{} = 4$, next-generation surveys must be able to map the $z \sim 2-3$ Ly$\alpha$ forest over a $\sim$10 times larger volume than fiducial \ztwonormal{} survey we have defined for Subaru PFS, i.e.\ $V_S \sim 3 \times 10^8\,\mpccubed$.
            Assuming a similar redshift range and target density as the PFS IGM tomography survey, this implies an survey area of $\sim 100\,\mathrm{deg}^2$, which is likely achievable in the next generation of large-scale spectroscopic surveys on large telescopes such as the Maunakea Spectroscopic Explorer \citep{MSE2016} or MegaMapper \citep{schlegel:2022} concepts.
            This also assumes the same magnitude limit and reconstruction quality as PFS, so this is a conservative estimate as future surveys will likely be able to improve on both avenues.
            In addition, as we find that projection from 3D onto 2D significantly degrades alignment significance, improved efforts to estimate 3D shape would provide immediate measured significance improvements.

\section{Conclusion} \label{sec:conclusion}

    In this paper, we analyzed the possible alignments between the large-scale ($\gtrsim$ Mpc) cosmic web and galaxy shapes/spins within the IllustrisTNG simulation at $z=1$ and $z=2$, with the goal of predicting
    whether they might be observable with the upcoming Galaxy Evolution Survey to be carried out on the Subaru PFS massively-multiplexed fiber spectrograph.
    
    We measure the relative 3D alignment between galaxy shapes and spins to the underlying cosmic web eigenvectors of the matter density field calculated using the pseudo-deformational tensor approach. 
    In the TNG300 simulation, above a transition stellar mass of $M_* > 10^{10.5}\,M_\odot$ we find that a clear excess alignment between galaxy shapes and the \ethree{} eigenvector, 
    which typically traces cosmic web filaments. 
    This alignment is present at comparable strengths at both $z=1$ and $z=2$, although the relative lack of high stellar mass galaxies at $z=2$  reduces the statistical significance of our finding at that earlier epoch. 
    The $M_* > 10^{10.2}\,M_\odot$ galaxy spins show a weak anti-alignment trend with respect to \ethree{} at $z=1$, 
    although the trend is difficult to detect clearly for lower-mass galaxies and at $z=2$.
    We find that these alignment trends are especially weak compared with the Horizon-AGN simulation, with Horizon-AGN alignment being $\sim5-6\sigma$ stronger than IllustrisTNG in the most conservative case.
    This is surprising given that Horizon-AGN is relatively contemporary with IllustrisTNG, indicating that galaxy-cosmic web alignment is a sensitive probe of the physics of galaxy formation and evolution.
    A detailed analysis of what process leads to this divergence is a promising future direction.
    We also compare with the smaller, but higher-resolution, TNG100 simulations where we generally find consistent alignment trends, although statistical uncertainties dominate in this smaller volume.
    
    We proceed to assess the observability of the this alignment signal in the Galaxy Evolution Survey planned to be carried out with the upcoming Subaru Prime Focus Spectrograph, 
    which will map the cosmic web with a high-number density galaxy redshift survey at $0.7 < z< 1.7$ as well as an IGM tomography survey of the Lyman-$\alpha$ forest at $2.2 < z< 2.7$. 
    We project the galaxy morphologies onto a 2D plane to mimic the observational shapes, 
    but within a volume the size of the TNG300 box we find that the projected alignment signal can vary depending on the direction from which the observer is viewing the data.
    We find that this viewing angle variance is a form of cosmic variance scaling with the observed volume, 
    which we attribute to anisotropic large-scale structures in the simulation volume being viewed from different directions. 
    This effect suggests even the $L=300$ cMpc TNG300 volume is not quite large enough to provide a fair sampling of the cosmic web alignments --- future theoretical studies should ideally use larger volumes with $L \gtrsim 1$ cGpc. 
    Equivalently, this additional cosmic variance should be taken into account in future observational analyses of galaxy alignments with the cosmic web, 
    although the cosmic volumes mapped by the Subaru PFS surveys will be $\sim 3-4\times$ the size of the TNG300 volume.
    
    Given the survey volume and number of galaxies expected to be observed in the lower-redshift ($z\sim1 $) PFS galaxies, we expect the shape alignment signal in TNG300 to be detectable at $5.3\,\sigma$ significance. 
    At $z\sim 2$, the signal will be more challenging to detect, rising only to $1.3\,\sigma$ significance above a random distribution. 
    This is based on the current baseline survey plan for $\sim 15,000$ coeval galaxies within the volume of the IGM tomographic map, 
    but a possible survey plan that doubles the number of coeval galaxies would only yield the expected $\sqrt{2}\sim 1.4\times$ boost.
    
    We estimate, however, that with the stronger alignment seen in previous analyses of the Horizon-AGN simulation, 
    there should a very strong alignment signal detectable in the PFS $z=1$ samples and even potentially at $z=2$. 
    
    However, the main obstacle to a real observational measurement of galaxy-cosmic web alignments on Subaru PFS is not the galaxy sample sizes nor cosmic web reconstruction, 
    but in the paucity of resolved near infrared imaging of galaxy morphologies within the proposed PFS footprint at the time of writing.
    The typical effective radii of quiescent galaxies ranges from $R_e \sim 1-2$ pkpc over $z\sim 1-2$, which correspond to $\theta \sim 0.2-0.3''$ on the sky. 
    Meanwhile, to cover the quiescent stellar distribution redward of the restframe $\lambda \approx 4000\,\AA$ break for $z> 1.5$ galaxies would require filter coverage at wavelengths beyond $1$ micron wavelengths, i.e.\ into the near-infrared. 
    This therefore calls for space-based $H$-band ($\lambda_C \sim 1.6\,\mu$m) imaging with the \textit{Hubble Space Telescope} (HST) to resolve stellar morphologies of galaxies up to $z\approx 2.7$. 
    Such imaging currently exists only over 1.4 deg$^2$ in the COSMOS field thanks to the recent HST-DASH survey, while another aggregate of $\sim 1$ deg$^2$ exists taking into account all other HST imaging. 
    This currently falls far short of the 12.3 deg$^2$ that the PFS GE survey will cover, although the Roman Space Telescope will efficiently allow this full footprint to be imaged once it is launched after 2029 --- some time after the completion of the Subaru PFS Galaxy Evolution Survey.
    
    However, once all the data is in place to allow constraints on the galaxy-cosmic web alignment signals at $z\sim 1-3$, one can expect this quantity to become an interesting probe of galaxy formation and evolution.  
 
\section{Acknowledgments}
We thank Viraj Pandya for useful discussions that helped initiate this project. 
Kavli IPMU was established by World Premier International Research Center Initiative (WPI), MEXT, Japan. K.G.L. acknowledges support from JSPS Kakenhi grants JP18H05868 and JP19K14755. 
B.H. is supported by the AI Accelerator program of the Schmidt Futures Foundation.
This research used resources of the National Energy Research Scientific Computing Center (NERSC), a U.S. Department of Energy Office of Science User Facility located at Lawrence Berkeley National Laboratory, operated under Contract No. DE-AC02-05CH11231.
The authors thank the Yukawa Institute for Theoretical Physics at Kyoto University. Discussions during the YITP workshop YITP-T-21-06 on ``Galaxy shape statistics and cosmology" were useful to complete this work.

\appendix

\section{TNG Shape-Spin Alignment}
\label{sec:appendix:shape-spin}

    \begin{figure*}[h]
        \centering
        \includegraphics[width=\figwidth]{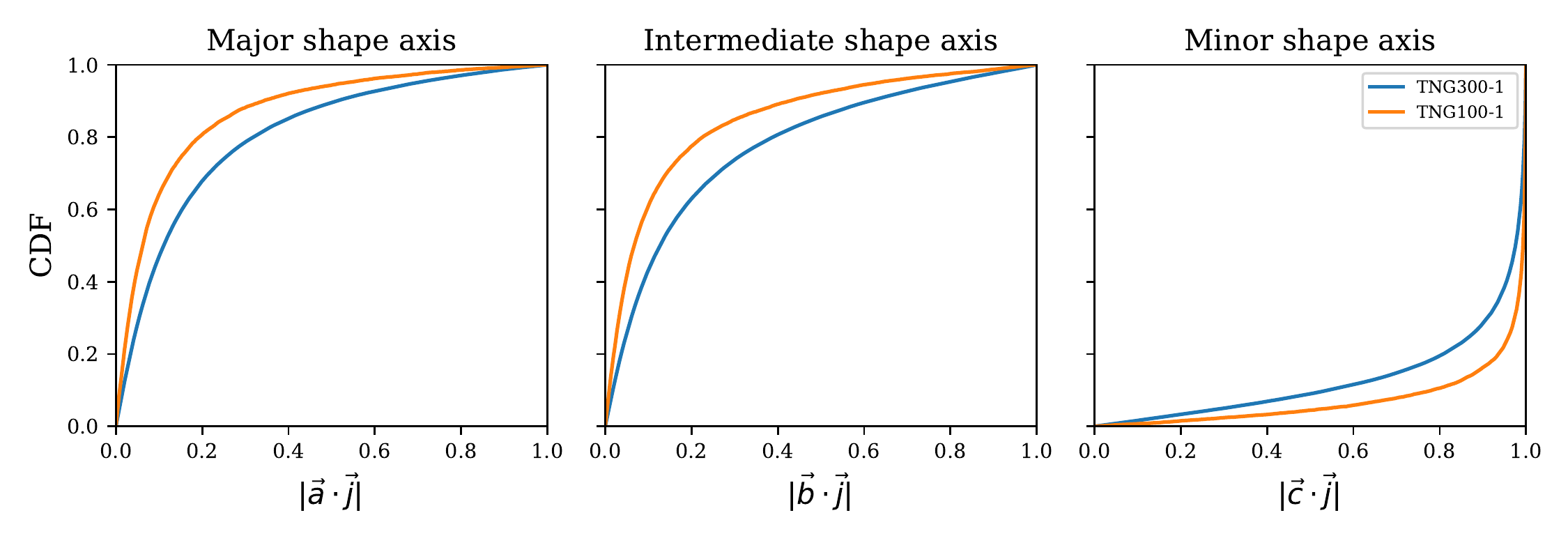}
        \caption{Cumulative density function of alignment between the spin and three shape ellipsoid axes of TNG300-1 and TNG100-1 galaxies, at \zone{}. The TNG galaxy sample is the same as Figure \ref{fig:Horizon-AGN-spin-comparison}, with a minimum-mass cut of $M_* > 10^{8.06}$ \msunh{} to match \citet{Codis2015Spin}.}
        \label{fig:shape-spin-misalign}
    \end{figure*}

    In Figure \ref{fig:shape-spin-misalign}, we plot the cumulative density function of alignment between the spin and three shape (inertia tensor) ellipsoid axes of TNG300-1 and TNG100-1 galaxies at \zone{}, without mock observational effects. 
    We use the same galaxy sample as Section \ref{sec:sim:results:horizon}, which aims to match the Horizon-AGN galaxy sample used in \citet{Codis2015Spin} by imposing a minimum stellar-mass cut of $M_* > 10^{8.06}$ \msunh{}.

    Comparing to the corresponding CDF plotted in Figure 2 of \citet{Codis2015Spin}, we see close qualitative agreement, with the minor shape axis closely aligned with the spin axis of galaxies.
    The slightly weaker degree of alignment between shape and spin in TNG300-1 compared to TNG100-1 may be due to worse mass resolution in the former simulation, as it persists across all mass bins.

    The high level of alignment between the minor shape axis and spin axis of galaxies suggests that the much weaker degree of spin alignment seen in TNG compared to Horizon-AGN extends to shape alignment.

\section{Alignment Dependence on Smoothing Scale}
\label{sec:appendix:smoothing}

    \begin{figure*}[h]
        \centering
        \includegraphics[width=0.45\textwidth]{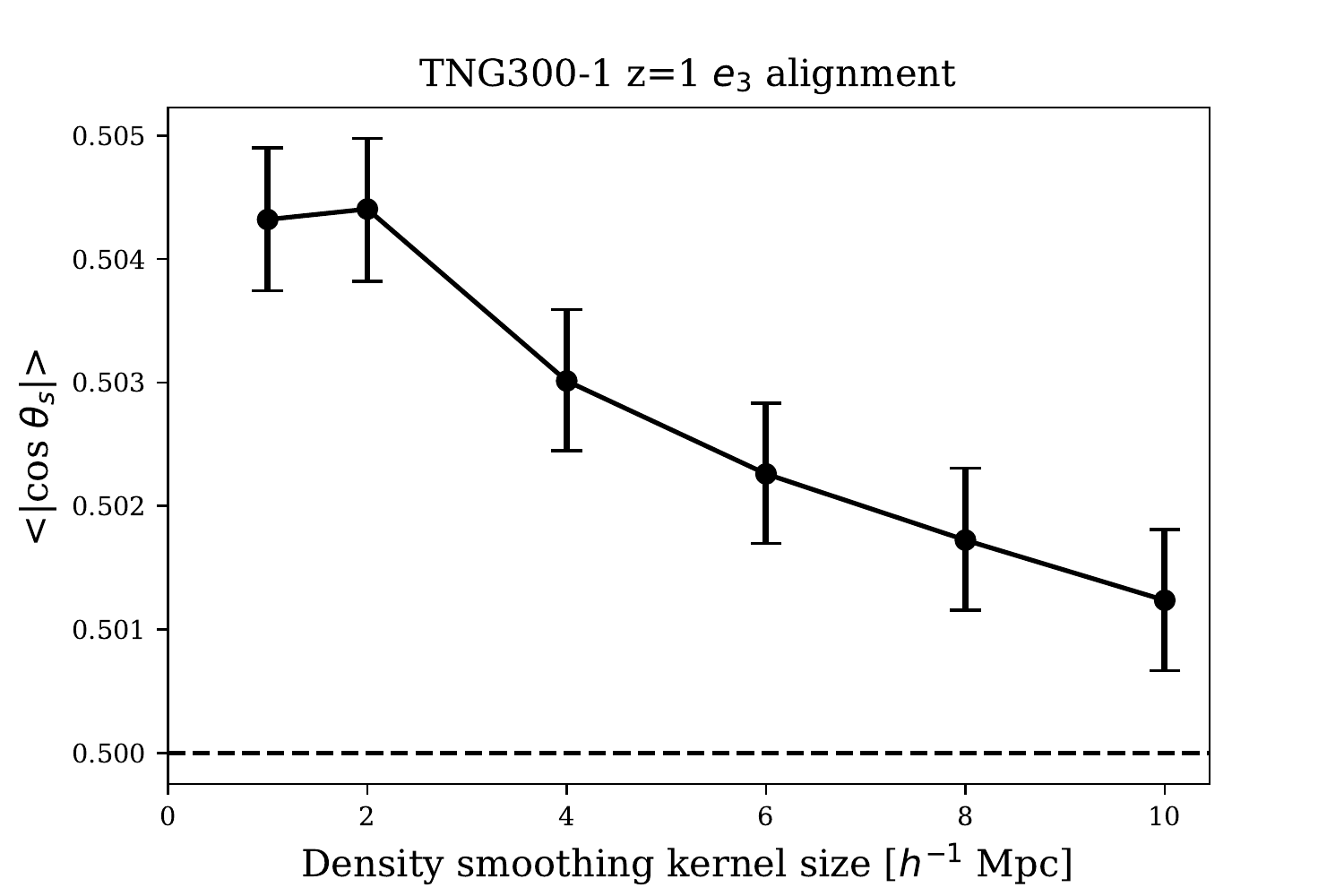}
        \caption{TNG300-1 $\ethree{}$-shape mean alignment, at \zone{}, as a function of density smoothing scale. We use the magnitude-selected \zonenormal{} galaxy sample, but do not apply any mock observational effects. Errors are derived from bootstrap resampling (but do not include any viewing-angle effects).}
        \label{fig:z1-shape-align-smoothing}
    \end{figure*}

    As cosmic web filaments exist on multiple scales, the degree of shape or spin alignment observed has a large dependence on the scale (Gaussian kernel size) at which the density is smoothed before calculating the deformation tensor.
    \citet{lee+20} find that as density smoothing scale increases, the magnitude of alignments decreases, which is corroborated in our analysis.
    In Figure \ref{fig:z1-shape-align-smoothing}, we plot shape-\ethree{} alignment as a function of density smoothing scale.
    We use the \zonenormal{} galaxy sample outlined in Section \ref{sec:mock:methods:gal-sample}, but do not apply any mock observational effects described in the section.

    Our choice of a 2 \mpc{} smoothing scale is motivated by the maximum reconstruction resolution we expect from \lya{} tomography; as described in \citet{lee_obs_req}, this is set by the average transverse \lya{} sightline spacing, which for PFS is $\sim 2$ \mpc{}.
    However, as Figure \ref{fig:z1-shape-align-smoothing} demonstrates, choosing a smoothing scale of $\sim 7$ \mpc{} would have cut the magnitude of alignment by a factor of two.
    We therefore recommend that analyses of alignment from PFS observational data adopt a cosmic web formalism that takes into account its multi-scale nature:
    for instance, by using the NEXUS cosmic web formalism \citep{cautun+13, caut14}, which extends the deformation tensor approach we use to multiple scales; or the Multiscale Morphology Filter \citep{ac07}, a similarly multi-scale algorithm which is based on the deformation tensor formalism.

\section{Mock Reconstruction Quality}
\label{sec:appendix:recon}

            \begin{figure*}[h]
                \centering
                \includegraphics[width=0.85\textwidth]{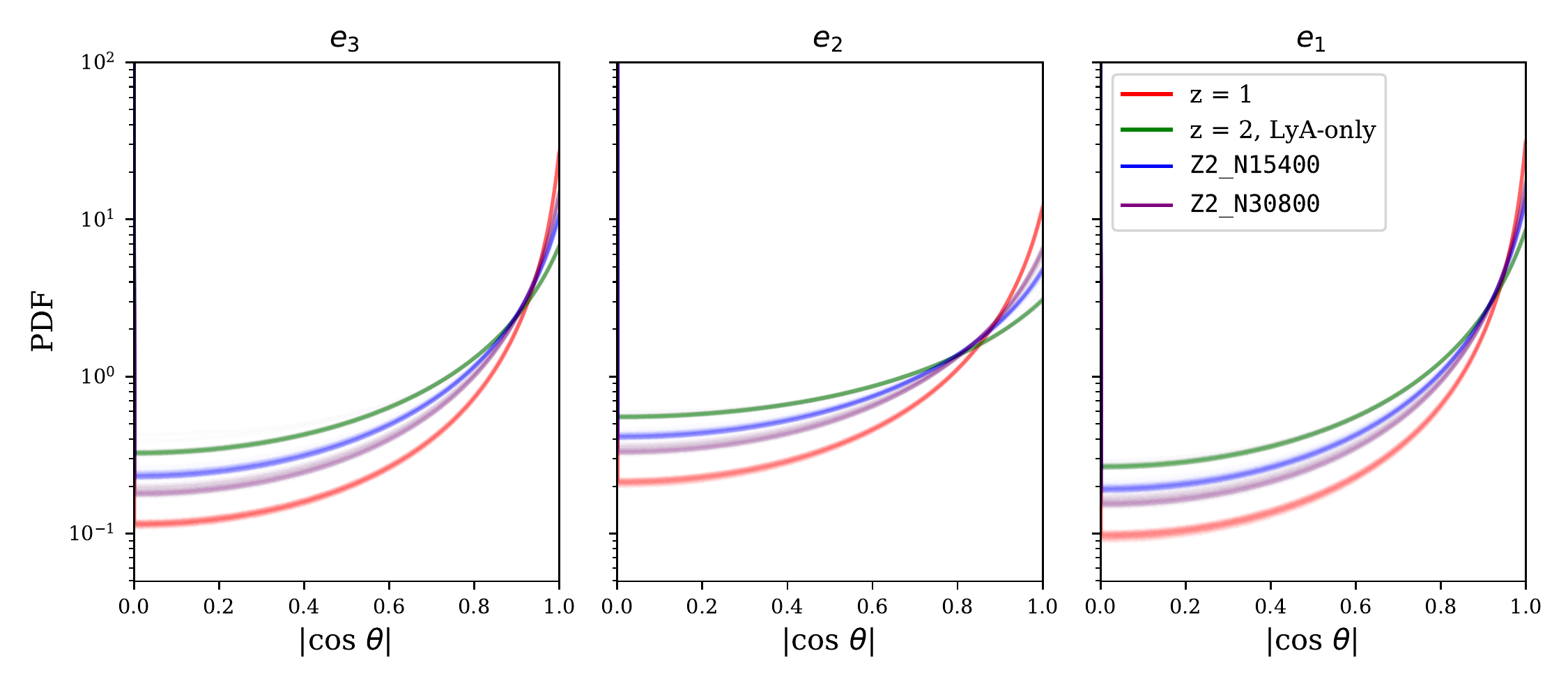}
                \caption{Probability density function of \alignment{} between true and \tardis{}-reconstructed eigenvectors, for each point in the reconstructed IllustrisTNG volume. Along with the \zone{} and joint \ztwo{} reconstructions used for the final analysis, we include a Lyman-alpha forest-only reconstruction at \ztwo{} for comparison. Each viewing angle's reconstruction is separately plotted, as a semi-transparent line. In the case of a perfect reconstruction, the PDF would be a delta-function at $\alignment{}  = 1$ for each of the panels.}
                \label{fig:recon-pdfs}
            \end{figure*}

            In this appendix, we validate the quality of the TARDIS density reconstructions on the galaxy redshift-space distribution of the \zonenormal{} sample at \zone{}, as well as the joint reconstructions of the galaxy distribution as well as Ly$\alpha$ forest absorption in the \ztwonormal{} and \ztwodouble{} mock samples at $z\sim 2$.
            For this purpose, we compute the eigenvectors of the pseudo-deformation tensor (as described in Section~\ref{sec:sim:methods:align}) for both the reconstructed volumes and true density field.
            Figure \ref{fig:recon-pdfs} illustrates the accuracy of the eigenvector recovery by plotting the PDF of \alignment{} between true and reconstructed eigenvectors at each grid point in the volumes, for all three of our mock observational survey configurations.
            We also include the performance of a \ztwo{} reconstruction using only Ly$\alpha$ tomography, to provide a baseline for the effect of incorporating the galaxy sample on reconstruction quality.
            In order to display variance in reconstruction quality across random viewing angles for a particular survey configuration, each viewing angle's PDF is plotted separately as a semi-transparent line.
            
            We find that for all survey configurations, \ethree{} and \eone{} are reconstructed more accurately than \etwo{}. This is consistent with results reported by \citet{TARDISII}.
            The \zone{} galaxy-only reconstruction is significantly more accurate than both \ztwo{} joint reconstructions, which are in turn both more accurate than the Lyman-$\alpha$ only reconstruction.
            Doubling the galaxy density for \ztwo{} yields a modest improvement, pushing the reconstruction quality closer to that of \zone{}.
            Overall reconstruction quality remains consistent across viewing angles for all survey configurations.
            However, we note that this does not imply that for a particular grid point in the volume, the reconstructed eigenvector will be as consistent.
            We find that for individual grid points, the reconstructed-true eigenvector alignment follows the same distribution as the overall reconstruction quality's, when comparing across viewing angles.

            Following the cosmic web classification scheme described in \cite{ForeroRomero2009TWeb} and \cite{LeeWhite2016} where the cosmic web is divided into nodes/sheets/filaments/voids using the eigenvalues of the deformation tensor, we also find that our classification accuracy is consistent with the accuracy reported by \citet{TARDISII}.

\bibliography{main}{}
\bibliographystyle{aasjournal}

\end{document}